\documentclass[apj]{emulateapj}
\usepackage[colorlinks,linkcolor={blue},citecolor={blue},urlcolor={blue}]{hyperref}
\usepackage{threeparttable}
\usepackage{bm}
\usepackage{graphicx}
\usepackage{epsf}
\usepackage{graphics}
\usepackage{amsmath}
\usepackage{lineno}
\usepackage{multirow,booktabs}

\def\beq{\begin{equation}}
\def\eeq{\end{equation}}
\def\bey{\begin{eqnarray}}
\def\eey{\end{eqnarray}}

\def\Mpc{\,{\rm Mpc}}
\def\mpc{\, h^{-1}{\rm {Mpc}}}

\def\Kpc{\, {\rm {kpc}}}

\def\kms{\,{\rm {km\, s^{-1}}}}
\def\Msun{{\rm M_\odot}}
\def\Zsun{{\rm Z_\odot}}
\def\msun{\, h^{-1}{\rm M_\odot}}
\def\msunt{\, h^{-2}{\rm M_\odot}}

\def\gs{\mathrel{\raise1.16pt\hbox{$>$}\kern-7.0pt
\lower3.06pt\hbox{{$\scriptstyle \sim$}}}}
\def\ls{\mathrel{\raise1.16pt\hbox{$<$}\kern-7.0pt
\lower3.06pt\hbox{{$\scriptstyle \sim$}}}}
\def\gtsima{\, {\buildrel > \over \sim} \,}
\def\ltsima{\, {\buildrel < \over \sim} \,}
\def\prosima{\, {\buildrel \propto \over \sim} \,}
\def\gsim{\lower.5ex\hbox{\gtsima}}
\def\lsim{\lower.5ex\hbox{\ltsima}}
\def\simgt{\lower.5ex\hbox{\gtsima}}
\def\simlt{\lower.5ex\hbox{\ltsima}}
\def\simpr{\lower.5ex\hbox{\prosima}}

\def\mstar{M_{*}}

\def\simba{\rm S{\scriptsize IMBA}}
\def\mufasa{\rm M{\scriptsize UFASA}}
\def\gizmo{\rm G{\scriptsize IZMO}}
\def\grackle{\rm G{\scriptsize RACKLE}}
\def\gadget{\rm GADGET-3}

\def\GZSBrw{\rm GZ-SBrw} 
\def\GZSBrs{\rm GZ-SBrs} 
\def\MSZR{$\mstar$ - $Z_{*}$ relation}
\def\MIZR{$\mstar$ - $Z_{\rm ISM}$ relation}
\def\fedd{f_{\rm Edd}}

\shorttitle{Coma galaxy cluster simulations}
\shortauthors{X., Luo et al.}

\begin{document}
\title{ELUCID VIII: Simulating the Coma Galaxy Cluster to Calibrate Model and Understand Feedback\vspace{-4em}}

\author{Xiong Luo\altaffilmark{1,2}, Huiyuan Wang\altaffilmark{1,2}, Weiguang Cui\altaffilmark{3,4,5}, Houjun Mo\altaffilmark{6}, RenJie Li\altaffilmark{1,2},  Yipeng Jing\altaffilmark{7,8}, Neal Katz\altaffilmark{6}, Romeel Dav\'{e}\altaffilmark{5,10,11}, Xiaohu Yang\altaffilmark{7,8,9}, Yangyao Chen\altaffilmark{1,2}, Hao Li\altaffilmark{1,2}, Shuiyao Huang\altaffilmark{6}}
\altaffiltext{1}{Key Laboratory for Research in Galaxies and Cosmology, Department of Astronomy, University of Science and Technology of China, Hefei, Anhui 230026, China; luoxiong@mail.ustc.edu.cn, whywang@ustc.edu.cn} 
\altaffiltext{2}{School of Astronomy and Space Science, University of Science and Technology of China, Hefei 230026, China}
\altaffiltext{3}{Departamento de F\'{i}sica Te\'{o}rica, Universidad Aut\'{o}noma de Madrid, M\'{o}dulo 15, E-28049 Madrid, Spain}
\altaffiltext{4}{Centro de Investigaci\'{o}n Avanzada en F\'isica Fundamental (CIAFF), Facultad de Ciencias, Universidad Aut\'{o}noma de Madrid, 28049 Madrid, Spain}
\altaffiltext{5}{Institute for Astronomy, University of Edinburgh, Royal Observatory, Edinburgh EH9 3HJ, United Kingdom}
\altaffiltext{6}{Department of Astronomy, University of Massachusetts, Amherst MA 01003-9305, USA}
\altaffiltext{7}{Department of Astronomy, School of Physics and Astronomy, Shanghai Jiao Tong University, Shanghai 200240, China}
\altaffiltext{8}{Tsung-Dao Lee Institute, Shanghai Jiao Tong University, Shanghai 200240, China}
\altaffiltext{9}{Shanghai Key Laboratory for Particle Physics and Cosmology, Shanghai Jiao Tong University, Shanghai 200240, China}
\altaffiltext{10}{University of the Western Cape, Bellville, Cape Town 7535, South Africa}
\altaffiltext{11}{South African Astronomical Observatories, Observatory, Cape Town 7925, South Africa}

\begin{abstract}
We conducted an investigation of the Coma cluster of galaxies by
running a series of constrained hydrodynamic simulations with \gizmo-\simba\ and \gadget\, 
based on initial conditions reconstructed from the SDSS survey volume in the ELUCID project. We compared simulation
predictions and observations for galaxies, ICM and IGM in and around the Coma cluster to constrain galaxy formation physics. 
Our results demonstrate that this type of constrained investigation allows us to probe in more detail
the implemented physical processes, because the comparison between simulations
and observations is free of cosmic variance and hence can be conducted in a ``one-to-one'' manner. 
We found that an increase in the earlier star formation rate and the supernova feedback of the
original \gizmo-\simba\ model is needed to match observational data on stellar, 
ISM and ICM metallicity. 
The simulations without AGN feedback can well reproduce the
observational ICM electron density, temperature, and entropy profiles, ICM substructures, and the IGM temperature-density relation, while the 
ones with AGN feedback usually fail.  However, one requires something like AGN feedback to reproduce a 
sufficiently large population of quiescent galaxies, particularly in low-density regions. 
The constrained simulations of the Coma cluster thus provide a test bed to understand processes that drive 
galaxy formation and evolution.
\end{abstract}


\section{Introduction} \label{secintro}

Observed galaxies are diverse in mass, color, morphology, size, star formation, nuclear activity, gas content, metallicity, 
and environment. These properties are usually related to each other, indicating that a complex network of 
processes drives the formation and evolution of galaxies. Various approaches have been used to understand these processes, 
and some key aspects have been identified, including the growth of the cosmic web and dark matter halos, 
gas cooling and condensation, ram pressure and tidal stripping, galaxy merger/interaction and instability, 
star formation and evolution, and feedback from stars and active galactic nuclei (AGNs) \citep[see e.g.][]{Mo2010}.
Many of the processes such as feedback, although originating from small scales, are expected to have significant impacts on the gas 
on larger scales, such as the intergalactic medium (IGM), the intracluster medium (ICM)
in clusters of galaxies, as well as the circumgalactic medium (CGM) \citep[][]{Cen2006ApJ,Fabian2012ARAA, heckman2014coevolution, Somerville2015physical, Tumlinson2017circumgalactic, nelson2019first, Eckert2021Univ, cui2021origin, Boselli2022, Yang2024}. 
Such a coupling between large and small scales makes it difficult to model the underlying 
processes accurately on the one hand, but provides the link between theoretical models and 
astronomical observations on the other.  

Cosmological hydrodynamic simulations can, in principle, implement many of the relevant processes. 
However, owing to limits on the numerical resolution and computational power,  some of the key processes can only 
be implemented as subgrid prescriptions with parameters which need to be calibrated by observational data and/or 
higher resolution simulations. These simulations have been quite successful in reproducing 
many properties of the gas and stellar components of the universe at different cosmic epochs, such as the 
stellar mass/luminosity functions, the star formation rates and gas content of galaxies, the stellar mass-supermassive 
black hole mass relation, and stellar mass-halo mass relation \citep[e.g.][]{Vogelsberger2014MNRAS, Hirschmann2014MNRAS, Dubois2014HorizonAGN, hopkins2014galaxies, Schaye2015MNRAS, Wang2015Nihao, Pillepich2018MNRAS, dave2019simba, cui2022three}. 
However, these apparently `successful' models usually make different assumptions for 
the same subgrid physics and adopt different prescriptions to treat the same physical processes, 
indicating that the details of these processes are still poorly understood and modeled. 

Clearly, to differentiate different hypotheses, realistic models and reliable calibrations of model parameters are needed \citep[e.g.][]{dave2019simba}. Traditionally, free parameters describing the subgrid physics 
are calibrated using small, high-resolution simulations. For example, the galaxy model for the Illustris simulations 
\citep{Vogelsberger2014MNRAS}, which the IllustrisTNG simulations \citep{weinberger2017simulating,Pillepich2018MNRAS,Springel2018,nelson2019IllustrisTNG} were also based on, 
was calibrated using simulations with a side length of 25$\mpc$ \citep{Vogelsberger2013}, and the EAGLE galaxy 
formation model was calibrated and tested using simulations with side lengths of 50 and 25$\mpc$, respectively
\citep{Crain2015EagleCalib}. On a scale of 25$\mpc$, the typical density fluctuation at $z=0$ is
already greater than 70\% \citep{Somerville2004, Chen2019CV}, indicating that cosmic variance
is a significant issue affecting the accuracy of the calibration. 
Cosmic variance also affects observational statistics used for the 
calibration of models. For example, \cite{Chen2019CV} found that the stellar mass function of SDSS galaxies 
is significantly underestimated 
for faint galaxies that are observed in relatively small volumes.

On the other hand, the potential of observational data has not yet been fully exploited 
in the model-data comparison. Statistical quantities of galaxies commonly used for model 
calibrations are the results of various mechanisms on various scales, and they alone may 
not be able to provide full information about the underlying processes. Indeed, 
to disentangle different processes it is necessary to compare observations and simulations in different 
mass components in different environments separately. For example, the dominant process driving the 
evolution of the gas on large scales ($\rm \gtrsim 1 Mpc$) is gravitational and can now be modeled accurately 
by numerical simulations. However, processes operating on smaller scales, in particular those related to feedback, 
which are difficult to simulate accurately, can also affect gas on large scales. Thus, the gas properties on large scales 
provide an important avenue to understand the underlying feedback processes. Unfortunately, 
few studies in the literature have directly used observations of the gas distribution on large scales to 
constrain galaxy formation models. Second, conventional simulations, which assume random phases
of Fourier modes in a relatively small simulation box, may not be able to match the 
ecosystems for galaxies in the observed volume, owing to cosmic variance as mentioned above. 
In contrast to traditional simulations, constrained simulations (CSs hereafter) are designed to reproduce the real 
large-scale structures found in a given volume of the universe, which allows simulated galaxies 
and structures to reside in the same environments and ecosystems as their observed counterparts, 
making the comparison between simulation results and observational data less affected by potential cosmic 
variance and hence more accurate.  For example, one may select a particularly interesting volume, where rich 
multi-band observational data are available, to carry out constrained simulations, and use comparisons 
between simulations and observations to calibrate models of galaxy formation. Since such ``one-to-one'' 
comparisons are, in principle, free of the cosmic variance, the results are thus much more powerful 
in probing the underlying physical processes. In addition, such constrained simulations 
also reconstruct the formation histories of the objects in question, which allows us to 
investigate the effects of history on the `remnants' we observe today. 

The initial condition of a constrained simulation is usually reconstructed from galaxy redshift surveys and/or 
peculiar velocity surveys, and many reconstruction methods have been developed 
\citep[e.g.][]{Hoffman1991ApJ, Nusser1992ApJ, Frisch2002Natur, Klypin2003ApJ, Jasche2013MNRAS, 
Kitaura2013MNRAS, WangH2013ApJ, WangH2014ApJ, Sorce2016MNRAS, Modi2018JCAP, Horowitz2019ApJ, Bos2019MNRAS,Kitaura2021MNRAS, LiY2022}. 
\citet{WangH2014ApJ} developed a method that implements particle-mesh (PM) dynamics 
to predict the final density field from a given initial density field, and uses
a Hamiltonian Monte Carlo Markov Chain to obtain the posterior model parameters 
(the set of Fourier modes representing the initial density field). Their tests based on N-body simulations 
show that the method can accurately recover the formation history of large-scale structures, 
even at scales of $\sim2\mpc$. This method has been successfully applied to the SDSS volume, 
as shown in \citet{WangH2016ApJ}. 

The Coma galaxy cluster is one of the most interesting structures in the local universe
suitable for our ``one-to-one'' study.  This cluster has attracted a lot of attention because 
it is one of the most massive, nearby galaxy clusters
with a mass of $M_{\rm 200c} \approx 6.2 \times 10^{14}\msun$ \citep{Okabe2014Subaru} and a redshift of 0.0241\citep{Yang2007ApJ}.
Large amounts of high-quality observational data have been accumulated for the Coma cluster, from 
the radio band all the way to the X-ray band 
\citep{Brown2011MNRAS, matsushita2011radial, petropoulou2012environmental, ogrean2013first, ade2013planck, 
akamatsu2013suzaku, Simionescu2013ApJ, Mirakhor2020MNRAS, Churazov2021A&A}. 
These observations can be used to understand the details of different mass components in and 
around the Coma cluster. For example, a radio relic around the virial radius, 
which is likely caused by a merger shock, has been detected by radio observations 
\citep{Brown2011MNRAS}; a strong bow-like shock in the inner region of the cluster 
has been revealed by X-ray observations \citep{Churazov2021A&A}; and 
optical observations show that the Coma cluster is connected to several massive 
filaments and contains two bright central galaxies (BCGs). All these provide 
important information about the formation and evolution of this particular 
cluster and about the underlying processes that drive the evolution of galaxies and gas components
in general.

Constrained simulations (CSs) have already been used to study the Coma cluster.
\citet{malavasi2023cosmic} used CSs to study the connection between the Coma cluster and the surrounding 
filaments, such as their spatial configuration, kinematic state, and evolution. 
\citet{Dolag2016MNRAS} studied the thermal Sunyaev-Zeldovich (SZ) effect in their simulated Coma cluster 
and found that the predicted Compton $y$-profile is slightly lower than that observed. Based on the ELUCID 
reconstruction \citep{WangH2014ApJ, WangH2016ApJ}, \citet{Li2022ELUCID} found that their 
predicted Compton $y$ and X-ray brightness profiles are in good agreement 
with the observed profiles within about half of the virial radius. 
In particular, their CS can reproduce bow-like shocks in the same locations 
as those revealed in observations, and these shocks are associated with recent mergers 
\citep[see also][]{Zhang2019MNRAS}. The CS used by \citet{Li2022ELUCID} does not include AGN feedback, 
commonly believed to have a significant impact on both galaxies and the ICM, and hence the problem needs to 
be revisited.

In this paper, we use initial conditions obtained from the ELUCID project \citep{WangH2014ApJ, WangH2016ApJ}
to carry out hydrodynamic simulations in a volume around the Coma cluster. These simulations 
implement different galaxy formation models, so that we can see directly how galaxy formation 
physics is reflected in the predicted galaxy population and gas components in different environments. 
This paper is organized as follows. In Section \ref{sec_obs}, we present observational data for the Coma cluster, 
including the galaxy catalog and the reduced data for the interstellar medium (ISM) and ICM. We also introduce an observational result for 
the IGM obtained in the SDSS region. In Section \ref{sec_sims}, we describe the initial conditions for the 
constrained simulations, the two simulation codes used, \gizmo-\simba\ and \gadget\, and the method 
to identify halos and galaxies. We compare the predictions of the fiducial models of the \gizmo-\simba\ and \gadget\
simulations with observational data in Section \ref{sec_fiducialmodelpred}. In Section \ref{sec_calimp}, we  
use the observational data for the Coma cluster to calibrate model parameters, and 
discuss the origin of the discrepancy between observations and simulations in connection to the role of 
AGN feedback. Finally, in Section~\ref{sec_sum}, we summarize our main results and make some 
further discussions. 

\section{Observational data}\label{sec_obs}

The galaxy sample we used is taken from the New York University Value-Added Galaxy Catalog 
\citep[NYU-VAGC,][]{Blanton2005NewYork}
based on SDSS DR7, which is also used for the ELUCID reconstruction (see below). 
The stellar masses of galaxies are obtained using the relation between the stellar mass-to-light ratio 
(M/L) and the color based on model magnitudes \citep[see][]{bell2003optical, Yang2007ApJ}.
The initial mass function (IMF) used to calculate the stellar mass is the Kroupa IMF
\citep{kroupa2001variation}, which differs from the Chabrier IMF \citep{chabrier2003galacticstellar} used in the simulations. 
However, both Kroupa and Chabrier IMFs have a stellar M/L that is about 0.25 dex lower than 
the Salpeter IMF \citep{salpeter1955luminosity}. Thus, no correction is needed 
when comparing simulated and observed galaxies, as discussed in \citet{borch2006stellar}. 

The specific star formation rate (sSFR), defined as the ratio between the star formation rate (SFR) of a galaxy and its stellar mass, is taken from \citet{brinchmann2004physical}, derived either directly from emission lines or indirectly from the 4000-\r{A} break. Star-forming and quiescent galaxies are separated by a value of $\log\rm sSFR=-11$, the same as that used for the simulations. The stellar metallicity, $Z_*$, of galaxies is taken from the eBOSS Firefly Value-Added Catalog \citep{comparat2017stellar} 
DR16\footnote{eBOSS Firefly Value-Added Catalog:  \url{https://www.sdss4.org/dr16/spectro/eboss-firefly-value-added-catalog/}}. It is measured assuming a Chabrier IMF and the MILES library \citep{sanchez2006medium}. In this paper, we take the solar metallicity $\Zsun$ to be the conventional value of 0.02, following previous studies \citep[e.g.][]{Borgne2003STELIB, Bruzual2003stellar,Ma2016origin}.
\cite{Tremonti2004} derived the gas-phase oxygen abundance 12+log(O/H) 
using optical nebular emission lines for star-forming galaxies in the SDSS. We use their data to compare with the ISM oxygen abundance measured in simulated galaxies. We take the gas-phase oxygen abundance as the metallicity of the ISM ($Z_{\rm ISM}$) hereafter.

We corrected the redshift distortion of these galaxies using the method shown in \cite{WangH2016ApJ}. This method,  
which employs linear theory to correct for the Kaiser effect \citep[e.g.][]{WangH2009MNRAS, WangH2012MNRAS} and uses the group 
catalog of \cite{Yang2007ApJ} to correct for the finger-of-God effect, has been shown to be capable of correcting for the 
anisotropy in the two-dimensional correlation function \citep[e.g.][]{Shi2016ApJ}.
Based on the corrected galaxy positions,  we select two galaxy samples according to their distances from 
the Coma cluster. The first sample of galaxies resides in a high-density region (HDR), located within $3R_{\rm 200c}$ (see the definition of $R_{\rm 200c}$ in section \ref{sec_galaxy_halo_def})
of the Coma cluster center, and the second sample contains galaxies in a low-density region (LDR) between $3R_{\rm 200c}$ 
and $7R_{\rm 200c}$ from the cluster center.  There are 135 (371) and 166 (94) star-forming (quiescent) galaxies 
with ${\rm log}(M_{*} / \msunt) > 9$ in the HDR and LDR, respectively.  
These counts are reduced to 43 and 107 for star-forming galaxies with available data for the oxygen abundance. 

In the following, we briefly describe the observational data of the ICM used in this paper. \cite{matsushita2011radial} analyzed 28 galaxy clusters, observed with XMM-Newton, including the Coma cluster, our subject of study in this paper. They integrated the spectra in 
each of the 6 concentric annular regions centered on the X-ray peak, and fitted the composite spectra with a multi-temperature model in the 0.5$-$10 keV band.  They then determined the Fe abundance from the flux ratios of Fe lines to the continuum within an energy range of 3.5$-$6 keV. Finally, they obtained the ICM Fe-abundance and temperature profiles for the Coma cluster with projected radius up to $0.5R_{180c}$. We correct their assumed solar abundance value of 0.0133 to our assumed value of 0.02.

\cite{Simionescu2013ApJ} presented X-ray data from Suzaku for the Coma cluster. The data consists of 24 pointings covering the E, NW, and SW directions contiguously out to a radius of $2^{\circ}$, combined with more pointings towards the NE and W from the cluster center. They modeled the hot gas as a thermal plasma with a single temperature in collisional ionization equilibrium within each shell. Furthermore, they performed a fit to the spectra in each annulus to measure the metallicity. They adopted a solar metallicity of 0.017, which we correct to 0.02. Both temperature and metallicity profiles obtained along azimuths not aligned with the infalling southwestern subcluster (i.e. E+NE+NW+W) are given in their paper. 
 With the assumption of spherical symmetry, 
they derived de-projected radial profiles of the electron density and specific entropy along the more relaxed NW+W 
and E+NE directions. In this paper, we only show their averaged profiles along the NW+W and E+NE directions. 

\cite{Mirakhor2020MNRAS} presented a new extended XMM-Newton mosaic of the Coma cluster that extends beyond its virial radius with almost complete azimuthal coverage. After combining it with the SZ observation from the {\it Planck}, 
they obtained the thermodynamic properties of the intracluster medium in an azimuthally averaged profile as well as in 36 angular sectors. They folded an APEC model through the XMM-Newton response in XSPEC and 
used the 0.7-1.2 keV count rate to derive the APEC normalization used to calculate the electron density. 
The derived electron density profile is deprojected by assuming spherical symmetry. They then derived the pressure profile 
$P(R)$ by fitting the projected $y$ profile to a universal pressure formula. The corresponding 3-d temperature and 
entropy profiles are calculated using $T(R)=P(R)/k_{\rm B}n_{e}(R)$ and $K(R)=P(R)/n_{e}^{5/3}(R)$, respectively.

We also use observational data for the IGM.  \cite{Lim2018MNRAS} derived the IGM temperature as a function of 
local mass density in the volume covered by SDSS DR7. They assumed a broken power-law relation between the electron 
pressure and the reconstructed mass density field \citep{WangH2009MNRAS, WangH2016ApJ} to predict the SZ Compton $y$ 
parameter map over the SDSS sky. They then performed a Monte Carlo Markov Chain to constrain the free parameters in the 
broken power-law to yield the best match to the observed $y$ map.  The temperature-density relation is then derived 
by assuming a mean cosmic baryon fraction.  Although the relation is obtained in the whole SDSS volume rather than 
the region around the Coma cluster, it is still interesting to compare with our simulation results, as we will show below.

\section{Simulations}\label{sec_sims}

\subsection{ELUCID reconstruction of initial conditions and zoom-in realizations}
\label{sec_ELUCID}

\begin{figure*}
   \centering
   \includegraphics[scale=0.43]{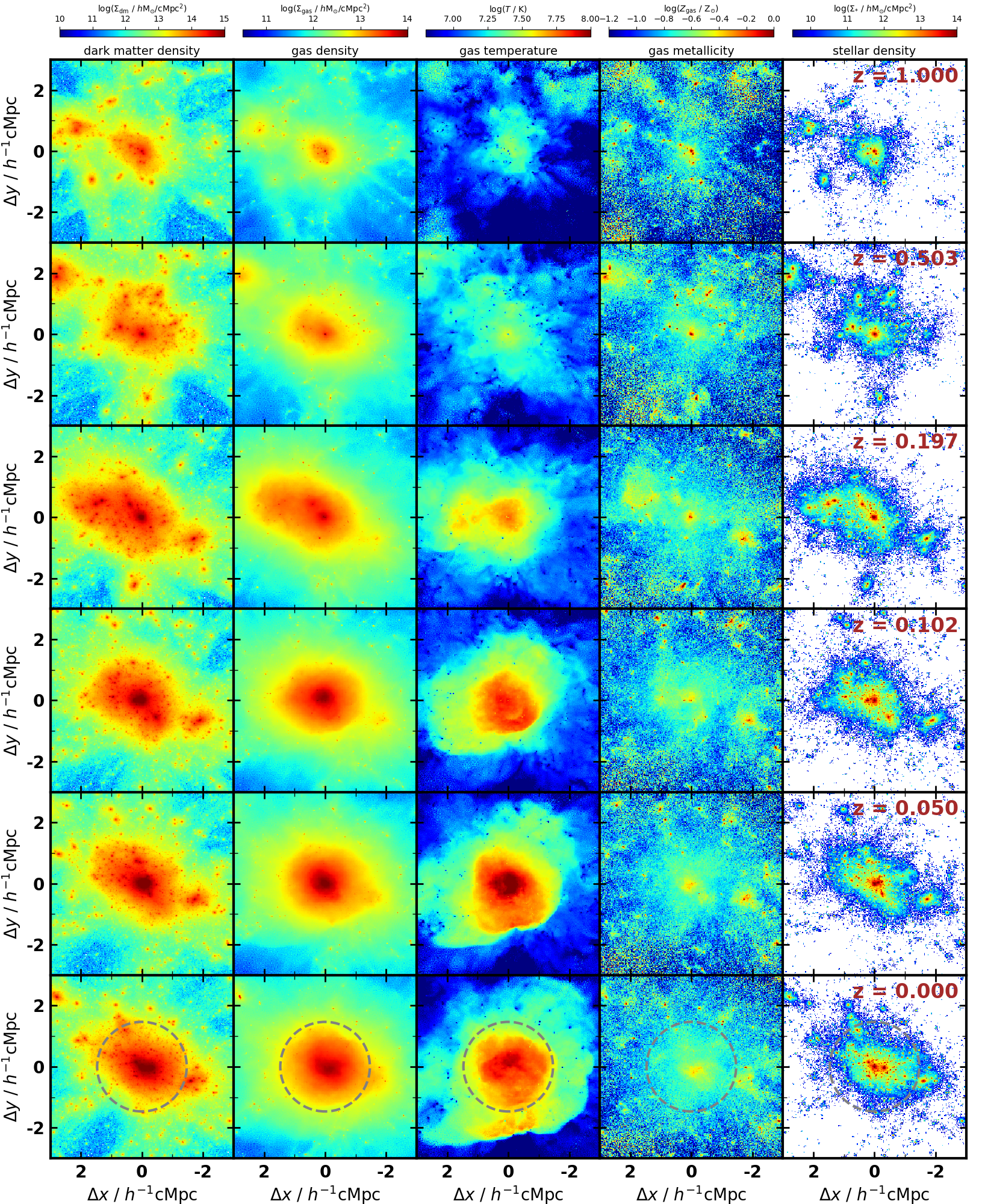}
	\caption{The evolution of the Coma cluster in the simulation \GZSBrw -CM from $z=1$ to 0. The columns from left to right 
 show dark matter density, gas density, gas temperature, gas metallicity and stellar density, respectively. The circles in the $z=0$ 
 panels indicate the virial radius of the cluster at the present time.}\label{fig_2Dmap}
\end{figure*}

The ELUCID project aims to reconstruct the initial conditions (ICs) of the local universe. The reconstruction method consists of four steps:
(i) identifying galaxy groups; (ii) correcting for redshift distortion; (iii) reconstructing the present-day density field, 
and (iv) recovering the ICs. A large constrained N-body simulation in the volume of the Sloan Digitial Sky Survey (SDSS) DR7 
\citep{Abazajian2009ApJS} up to $z = 0.12$ has been carried out, as shown in \cite{WangH2016ApJ}. In what follows, we refer 
to this constrained simulation as the original CS (hereafter OCS). We refer the reader to \cite{WangH2016ApJ} and references 
therein for details of the method and the OCS. In this paper, we focus mainly on one particularly interesting region, 
centered on the Coma galaxy cluster at a redshift of 0.0241. We adopt a zoom-in technique, which simulates structures of 
interest with a higher resolution than other regions in the simulation volume \citep{Katz1993ApJ}.

Details of the method to generate the zoom-in ICs are presented in \cite{Li2022ELUCID}. Here we present a brief description.  
First, we select a high-resolution region (HIR) containing the Coma galaxy cluster from the $z=0$ snapshot of the OCS. 
The HIR is chosen to be a spherical volume centered on the simulated Coma.  We follow all the dark matter particles 
in the HIR back to the initial time and select an initial HIR (iHIR) that contains all the HIR particles. 
The iHIR has the same comoving center and geometry as the HIR. To prevent lower-resolution particles from 
entering the HIR, we set up a buffer region with a size equal to 5\% of the iHIR. The cosmic density field 
outside the iHIR is sampled with particles with three different levels of lower resolution. Gas particles are 
only placed in the iHIR. We split each high-resolution particle into two,  
one dark matter particle and one gas particle with a mass ratio equal to 
$(\Omega_{\rm m,0}-\Omega_{\rm b,0})/\Omega_{\rm b,0}$, and separate the two particles 
in a random direction by a distance equal to half of the mean particle separation.

We adopt the same cosmological parameters \citep[WMAP5, ][]{Dunkley2009ApJS} as the OCS: $\Omega_{\Lambda,0}$ = 0.742, 
$\Omega_{m,0}$ = 0.258, $\Omega_{b,0}$ = 0.044, $h$ = $\rm H_0/(100\kms\Mpc^{-1})$ = 0.72, $\sigma_{8}$ = 0.80. The mass resolution of the HIR is 8 times as high as the OCS simulation. The corresponding 
masses of dark matter and gas particles in the iHIR are $3.20\times10^7\msun$ and $6.58\times10^6\msun$, respectively.  
The radii of the spherical volume of these simulations are set to be either 20 or 30 $\mpc$, much larger than the 
virial radius of the Coma cluster (see below). This enables us to study galaxies in both high- and low-density 
environments around the Coma cluster. In addition, we also have CSs for a void in the SDSS region at $z \approx 0.05$ to 
check cosmic variance when necessary. The HIR for the void simulation is a spherical volume with a radius of 
40$\mpc$. See \cite{Li2022ELUCID} for a detailed definition of the void.

Information about the simulations is presented in Table \ref{TableSimulations}. The name of each simulation
consists of three parts. The first part indicates the code used, either \gizmo (GZ) or \gadget\ (G3). 
The second part denotes the galaxy formation model, which is discussed in the following subsections. 
The third part specifies the simulated region, either the Coma galaxy cluster (CM) or the void (VD). 
For example, GZ-SB-CM is a constrained simulation of the Coma cluster run with \gizmo\ and \simba\ physics 
\citep{dave2019simba} (see Section \ref{sec_GZ-SM}), and G3-H-VD is a constrained simulation of the void with \gadget\ 
and the galaxy formation model presented in \cite{Huang2020MNRAS} (see Section \ref{sec_G3-H}). The first two parts 
of the name can also be used to indicate the model used. For example, the GZ-SBnA model is the \gizmo-\simba\ model 
with the AGN feedback switched off, and the G3-H model is the model presented in \cite{Huang2020MNRAS}.  These simulations are run with the support of the Jiutian simulation project.

\subsection{Simulation codes}

\subsubsection{\gizmo-\simba}\label{sec_GZ-SM}

\simba\ \citep{dave2019simba} is a successor to the earlier \mufasa\ simulations \citep{dave_2016}, both 
were developed on top of a branched version of the gravity plus hydrodynamics code \gizmo\ \citep{hopkins2015new} in its 
Meshless Finite Mass (MFM) mode. \simba's default setup, which was used to run the $100\ \mpc$ boxsize cosmological 
simulation in \citet{dave2019simba}, is strictly adopted here and named GZ-SB. Here, we provide brief descriptions 
of this fiducial model together with our modified models. They are also listed in Table \ref{TableSimulations} for reference.

This fiducial model includes both photo-ionisation heating and radiative cooling, using the \grackle-3.2 
library \citep{smith2017grackle} in its non-equilibrium mode. 
The neutral hydrogen fraction, with self-shielding following the \cite{rahmati2013evolution} prescription, 
is computed self-consistently within \grackle, with the metagalactic ionising flux attenuated according to  
the gas density. Furthermore, the ionising background is assumed to be uniform and to follow 
the \cite{haardt2012radiative} model.
A ${\rm H}_2$-based \cite{schmidt1959rate} law is used to model the star formation: 
${\rm SFR} = 0.02\rho_{\rm H_2}/t_{\rm dyn}$, where the H$_2$ density is computed from
the metallicity and local column density using the sub-grid prescription of 
\cite{krumholz2011comparison}. 

Although the newer SIMBA-C \citep{hough2023simba} adopts the advanced Kobayashi chemical model to track 34 different 
elements, we stick to the original chemical enrichment model in \simba\, which tracks 11 elements 
(H, He,  C, N, O, Ne, Mg, Si, S, Ca, Fe) from Type II supernovae (SNII), Type Ia supernovae (SN1a) and asymptotic 
giant branch (AGB) stars, as described in \cite{Oppenheimer2006MNRAS}. Yield tables for SNII from 
\cite{nomoto2006nucleosynthesis}, for SN1a from \cite{iwamoto1999nucleosynthesis} with 1.4$\Msun$ of metals per 
supernova, and for AGB from \cite{Oppenheimer2006MNRAS} (assuming a helium fraction of 36\% and a 
\cite{chabrier2003galacticstellar} initial mass function), are adopted in the simulation. 
The dust model, as described in \cite{li2019dust}, broadly follows that given in \cite{mckinnon2016dust} with the 
metal mass passively advected along with the gas. The growth of dust particle mass is through condensation, and 
the accretion of gas-phase metals is via two-body collisions. The dust inside a gas particle can be destroyed 
by supernova shocks, AGN feedback and collisions with thermally excited gas elements, and its mass and metals are 
then returned to the gaseous phase.

The stellar feedback is prescribed by two-phase, decoupled and metal-enriched winds, with 30\% of the wind particles 
ejected hot and the rest ejected at $T\approx 10^3$K.  The mass loading factor scales with the stellar mass of the galaxy 
based on the mass outflow rates from the Feedback In Realistic Environments \citep[FIRE,][]{hopkins2014galaxies} 
simulations \citep{angles-alcazar2017cosmic}, where the stellar mass is provided by an on-the-fly friends-of-friends 
galaxy finder. The wind velocity scaling is also based on results from the FIRE simulations \citep{Muratov2015MNRAS}, 
with modifications adopted in the \simba\ simulation. Ejected wind particles are temporarily hydrodynamically decoupled to 
avoid numerical inaccuracies. The cooling and other baryon processes are also paused during the decoupled time. 
Wind particles are enriched with metals from their surroundings at the launch time according to 
\cite{nomoto2006nucleosynthesis}. The star formation-driven wind mechanisms are described in more 
detail in \cite{appleby2021low}. 

\simba\ employs a novel method for supermassive black hole (SMBH) growth and feedback. SMBHs are seeded with 
$M_{\rm seed} \sim 10^4 \Msun$ into galaxies with $\mstar \gtrsim 10^{9.5} \Msun$ in the \simba\ simulation. 
However, the iHIR has a slightly higher resolution than  \simba. Therefore, we seed an SMBH particle when the host galaxy 
stellar mass is larger than $10^{9} \Msun$ by scaling the SMBH seeding mass to $\sim 3\times 10^3 \Msun$. 
The SMBH mass growth is through a two-mode accretion prescription in the simulation: For cold gas ($T< 10^5$\ K) surrounding 
the SMBH, torque-limited accretion is implemented following the prescription of 
\cite{angles-alcazar2017gravitational}, which is based on \cite{hopkins2011analytic}; For hot gas 
(T $> 10^5$\ K), classical Bondi accretion is adopted.

Kinetic AGN feedback, similar to stellar feedback, is implemented in two modes: ``radiative'' and ``jet'' at high 
and low Eddington ratios separated by $\fedd = 0.2$. This is designed to mimic the dichotomy in real AGN 
\citep[e.g.][]{heckman2014coevolution}. The radiative and jet feedback outflows are ejected in a $\pm$ direction along the angular 
momentum vector of the inner disk with zero opening angle and different velocities. The mass outflow rate varies so that 
the momentum output of the SMBH is $20L/c$, where $L$ is the bolometric AGN luminosity. In the radiative mode, 
winds are ejected at the ISM temperature with an SMBH-mass-dependent velocity calibrated using X-ray detected 
AGN in SDSS \citep{perna2017X-ray}:
\begin{equation}
    v_{\rm rad} = 500 + 500(\log M_{\rm BH} - 6)/3 \ {\rm km\ s^{-1}}.
\end{equation} 
At $\fedd < 0.2$ the transition to the jet mode begins for galaxies with $M_{\rm BH} > 10^{7.5}\Msun$, 
with winds particles ejected at the virial temperature of the halo. Jet outflows receive a velocity boost 
that increases with decreasing $\fedd$, up to a maximum of 7000 km s$^{-1}$ above the radiative mode 
velocity at $\fedd = 0.02$:
\begin{equation}
    v_{\rm jet} = v_{\rm rad} + 7000 \log (0.2/f_{\rm Edd})\ {\rm km\ s^{-1}}.
\end{equation}
In addition, \simba\ includes an X-ray mode of the AGN feedback, following \cite{choi2012radiative}, to mimic X-ray 
heating from the accretion disk. This X-ray feedback model is only turned on when the host galaxies have 
$f_{\rm gas} < 0.2$ and the AGN feedback is in the full-velocity jet mode. As shown in \cite{cui2021origin}, 
jet-mode feedback is the key to quenching galaxies while the X-ray feedback enhances 
color bimodality of the galaxy population.

In addition to the model GZ-SB that adopts the original \simba\ setup, we have other two models, \GZSBrw\ and \GZSBrs , based 
on the \simba\ simulation.  In Section \ref{sec_calib}, we describe the details of the two models.  In general, both 
star formation and supernova feedback are enhanced in the two models, while the former tries to shift the onset of the 
jet mode to higher redshift and the latter tries to simply increase the jet velocity. Moreover, we ran a simulation with 
the fiducial setup but without the SMBH part. This simulation, referred to as GZ-SBnA, is used to compare with 
the \gadget\ runs that also neglect the SMBH growth and AGN feedback.

\begin{table*}
    \centering
    \caption{A brief summary of the simulation set}
\begin{threeparttable}
    \begin{tabular}{cccccc}
    \toprule
    Simulation& Structure\tnote{(a)}& HIR Radius\tnote{(b)} & Code & SF model & AGN model \\
    \hline
    GZ-SB-CM & Coma& 20& \gizmo\ & \simba\tnote{(c)}& \simba\tnote{(c)} \\
    \GZSBrw -CM & Coma&  20 & \gizmo& high SF \& strong SN\tnote{(d)} & weak Jet\tnote{(e)} \\
    \GZSBrs -CM & Coma& 20 & \gizmo& high SF \& strong SN & strong Jet\tnote{(f)}\\
    GZ-SBnA-CM & Coma& 20 & \gizmo& \simba & no AGN \\
    \hline
    G3-H-CM & Coma& 30 & \gadget & Huang et al.\tnote{(g)} & no AGN\\
    G3-H-VD & Void& 40 & \gadget & Huang et al. & no AGN\\
    \bottomrule
    \end{tabular}
    
    \begin{tablenotes}
    \footnotesize
    \item[(a)] The name of the large-scale structure that we simulate. The Coma galaxy cluster is located at $\alpha_{\rm J}=194.8$, $\delta_{\rm J}=27.9$ and $z=0.0241$, and the mean location of the void is $\alpha_{\rm J}=222.6$, $\delta_{\rm J}=45.5$, and $z=0.0519$.
    \item[(b)] The HIR radius is in units of $\mpc$. 
    \item[(c)] This is the fiducial \gizmo-\simba\ model \citep{dave2019simba}.
    \item[(d)] Both star formation and supernova feedback are enhanced.
    \item[(e)] The jet velocity is similar to GZ-SB-CM, but we allow the jet at a lower velocity to heat the surrounding gas to the virial temperature to shift the jet mode-on time to higher redshifts.
    \item[(f)] The jet velocity is almost doubled to increase the AGN feedback strength.
    \item[(g)] The star formation and stellar feedback model is presented in \cite{Huang2020MNRAS}
    \end{tablenotes}
\end{threeparttable}
\label{TableSimulations}
\end{table*}

\subsubsection{\gadget}\label{sec_G3-H}

We also include simulations evolved using the \gadget\ code and the galaxy formation model presented in 
\cite{Huang2019MNRAS, Huang2020MNRAS}. We use names starting with ``G3-H'' to denote these simulations, 
as shown in Table~\ref{TableSimulations}. The G3-H-CM and G3-H-VD simulations are exactly those presented 
in \cite{Li2022ELUCID}. Unlike GZ-SB and its other versions, the G3-H model adopts an SPH technique to solve
fluid equations and a galaxy formation model without AGN feedback. Therefore, comparing G3-H, GZ-SB with observations can shed light on the importance of AGN feedback. We refer readers to \cite{Huang2019MNRAS}, \cite{Huang2020MNRAS} 
and \cite{Li2022ELUCID} for details of the code and the simulations. Here we only give a brief description. 

The \gadget\ code \citep{Huang2019MNRAS,Huang2020MNRAS}, an updated version of GADGET-2 \citep{Springel2005MNRAS},  
includes several numerical improvements in the SPH technique, such as using the pressure-entropy formulation
\citep{Hopkins2013MNRAS} to integrate fluid equations, using a quintic spline kernel over 128 neighbouring 
particles to measure fluid quantities, and adopting the \citet{Cullen2010MNRAS} viscosity algorithm and 
artificial conduction to capture shocks. These updates lead to considerable
improvements in the instabilities at fluid interfaces in subsonic flows 
\citep{Sembolini2016a, Sembolini2016b, Huang2019MNRAS}.

The radiative cooling of the G3-H model includes cooling from hydrogen, helium, and metal lines for altogether 11 elements \citep{Wiersma2009MNRAS}. The cooling rate is recalculated according to a uniform ionizing
background given by \citet{haardt2012radiative}. Following \citet{Springel2003MNRAS}, we use a subgrid approach to model the multiphase ISM in dense regions with $n_\mathrm{H} > 0.13\ \mathrm{cm^{-3}}$ and a star 
formation recipe that matches the Kennicutt-Schmidt law. The G3-H model traces the enrichment of four metal 
species, C, O, Si and Fe, produced from SNII, SN1a and AGB stars \citep{Oppenheimer2008MNRAS}.

We adopt a kinetic sub-grid model for the stellar feedback \citep[see also][]{Huang2019MNRAS, Huang2020MNRAS}.  
A SPH particle in star forming regions can be ejected as a wind particle with a probability proportional to 
the local star formation rate. We adopt the same set of wind parameters as the fiducial simulation 
from \cite{Huang2020MNRAS}, which matches a broad range of observations \citep[e.g.][]{Dave2013MNRAS}. 
This results in momentum-driven wind scalings for large $\sigma_{\rm gal}$ and supernova-energy-driven wind 
scalings for small $\sigma_{\rm gal}$,  
where $\sigma_{\rm gal}$ is the velocity dispersion of the galaxy. These scalings are very similar to those found in very high-resolution 
zoom simulations \citep{Hopkins2012MNRAS, hopkins2014galaxies,Muratov2015MNRAS}. We model the launch of galactic winds from star-forming galaxies with the mass loading 
factor, $\eta\equiv {\rm ejection~rate}/{\rm SFR}$. So defined we assume that, 
$\eta\sim\sigma_{\rm gal}^{-\beta}$ for small $\sigma_{\rm gal}$ and $\eta\sim\sigma_{\rm gal}^{-\alpha}$ 
for large $\sigma_{\rm gal}$. The initial wind speed, $v_{w}$, scales linearly with  $\sigma_{\rm gal}$. The wind scalings are assumed 
at the wind launch from the star-forming regions of the galaxy while the very high-resolution zoom simulations 
report their wind scalings at $0.25 r_{\rm vir}$ \citep{Muratov2015MNRAS}. Hence, we have had to slightly 
increase our wind launch velocities to reproduce their behavior at $0.25 r_{\rm vir}$. 
We also cap the wind speed so that the energy in the winds does not exceed that available in supernovae. 

\subsection{Galaxies and halos around the Coma cluster}
\label{sec_galaxy_halo_def}
Galaxies are identified using the Spline Kernel Interpolative Denmax (SKID) algorithm \citep[][]{Keres2005MNRAS}. 
This technique groups particles on the density gradient lines that converge on the same local density maxima. 
To do this, a smoothed density field is first constructed using all particles. Then, test particles at the 
position of star-forming gas and stars are moved along the positive gradient direction of the density field 
until a local density maxima is reached. All particles with test particles at the same local density maxima 
are considered as potential members of one galaxy. An unbinding procedure is then applied to the group of 
particles to determine the final member particles of the galaxy. The center of the galaxy is defined as the 
location of its density maxima. The star and/or star-forming gas particles in each galaxy can be used to 
calculate the stellar mass $M_{*}$, stellar metallicity $Z_{*}$, specific star formation rate (sSFR) and 
ISM oxygen abundance 12 + log(O/H). We divided the simulated galaxies into star-forming and quiescent galaxies 
using $\log\rm sSFR=-11$.

We use a FoF algorithm to identify halos. High-resolution dark matter particles are linked to each other within 
a link length $b$ equal to 0.2 times the mean separation of dark matter particles. Gas, star and black hole particles 
are also linked to their closest dark matter particle if their distances from the particle are less than $b$.  
The halo center is defined as the position of the dark matter particle that possesses the minimum potential 
among all the FoF particles. The halo virial radius, $R_{\rm 200c}$, is the radius of a sphere within which the 
average density is 200 times the critical density of the universe at that redshift, and the halo mass, $M_{\rm 200c}$, refers to the total mass within $R_{\rm 200c}$.  

The Coma clusters in these simulations have almost the same $M_{\rm 200c}$ and assembly history.  At $z=0$, the Coma cluster 
has $M_{\rm 200c} \approx 7.52 \times 10^{14}\msun$ and $R_{\rm 200c} \approx 1.48\mpc$, corresponding to $70.9'$ viewed from the earth.  
These are consistent with those of the real Coma cluster,  
$M_{\rm 200c} \approx 6.2 \times 10^{14}\msun$ \citep{Okabe2014Subaru} and  
$R_{\rm 200c} \approx 70'$ \citep{Simionescu2013ApJ}.  As an example, 
Figure \ref{fig_2Dmap} shows the evolution of dark matter density, gas density, gas temperature, gas metallicity, 
and stellar density around the Coma cluster in the \GZSBrw -CM simulation.  During the last 2.4 Gigayears, the Coma cluster 
experienced several massive merger events.  These mergers generate multiple shocks that heat the ICM and spread 
far beyond the virial radius, as can be seen in the temperature map.  Interestingly, our simulated Coma cluster
contains two BCGs residing in the center of the cluster, similar to the observed Coma cluster.

\begin{figure*}
   \centering
   \includegraphics[scale=0.41]{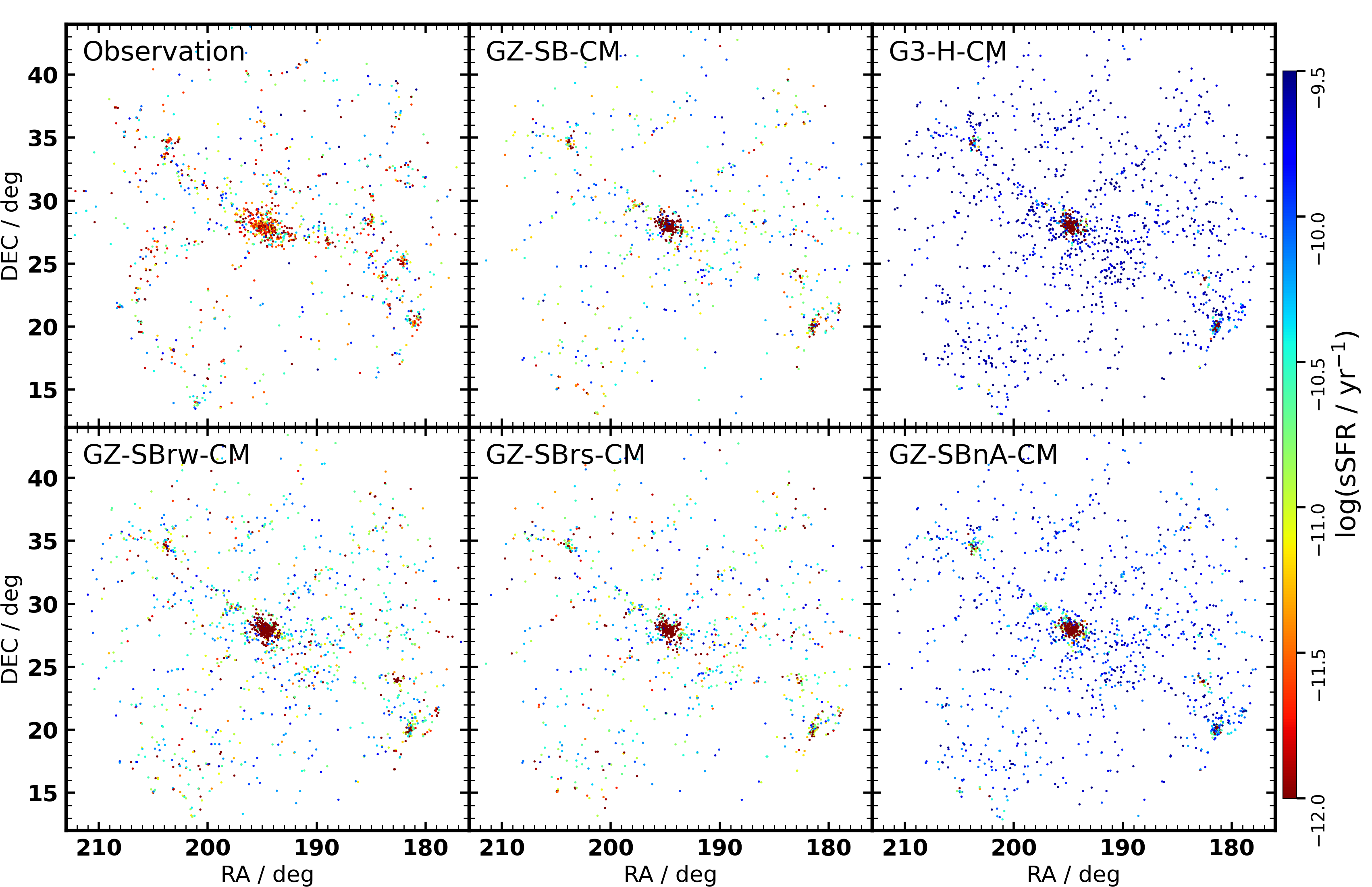}
	\caption{Two-dimensional distributions of galaxies in the J2000.0 coordinate system. 
        The horizontal and vertical axes are, respectively, the right ascension and declination in degrees. 
        The six panels show results from observations and five simulations as labeled. Only galaxies with $\log(M_*/\msunt)>9.5$ within a radius of 20 $\mpc$ around the Coma cluster are displayed, colored by their specific star formation rate.}\label{fig_gal2Dmap}
\end{figure*}

Figure \ref{fig_gal2Dmap} shows the spatial distributions (in J2000.0 coordinates) of the simulated and 
observed galaxies, as seen from the Earth. Galaxies with $\log(M_*/\msunt)>9.5$ are complete at this redshift 
(see below), and thus only these galaxies are shown. As one can see, the simulations reproduce the observed 
large-scale structures, such as the Coma cluster in the center, two smaller clusters at the southwest and 
northeast of the Coma cluster, and the filaments connecting the clusters. However, smaller structures are not 
accurately reproduced, as expected. One of the most significant discrepancies between the simulations and observation 
is the quiescent galaxy population in the low-density region. It appears that simulations produce too few quiescent galaxies, 
as we will discuss in the following section.

\section{Fiducial simulations versus observations}\label{sec_fiducialmodelpred}

In this section, we compare the predictions of the two fiducial simulations, GZ-SB-CM and G3-H-CM, with 
observational data. We focus on galaxies in Section \ref{sec_galpro},  and on the ICM and IGM in Section \ref{sec_ICMIGM}.

\subsection{Properties of galaxies}
\label{sec_galpro}

\begin{figure*}
   \centering
   \includegraphics[scale=0.41]{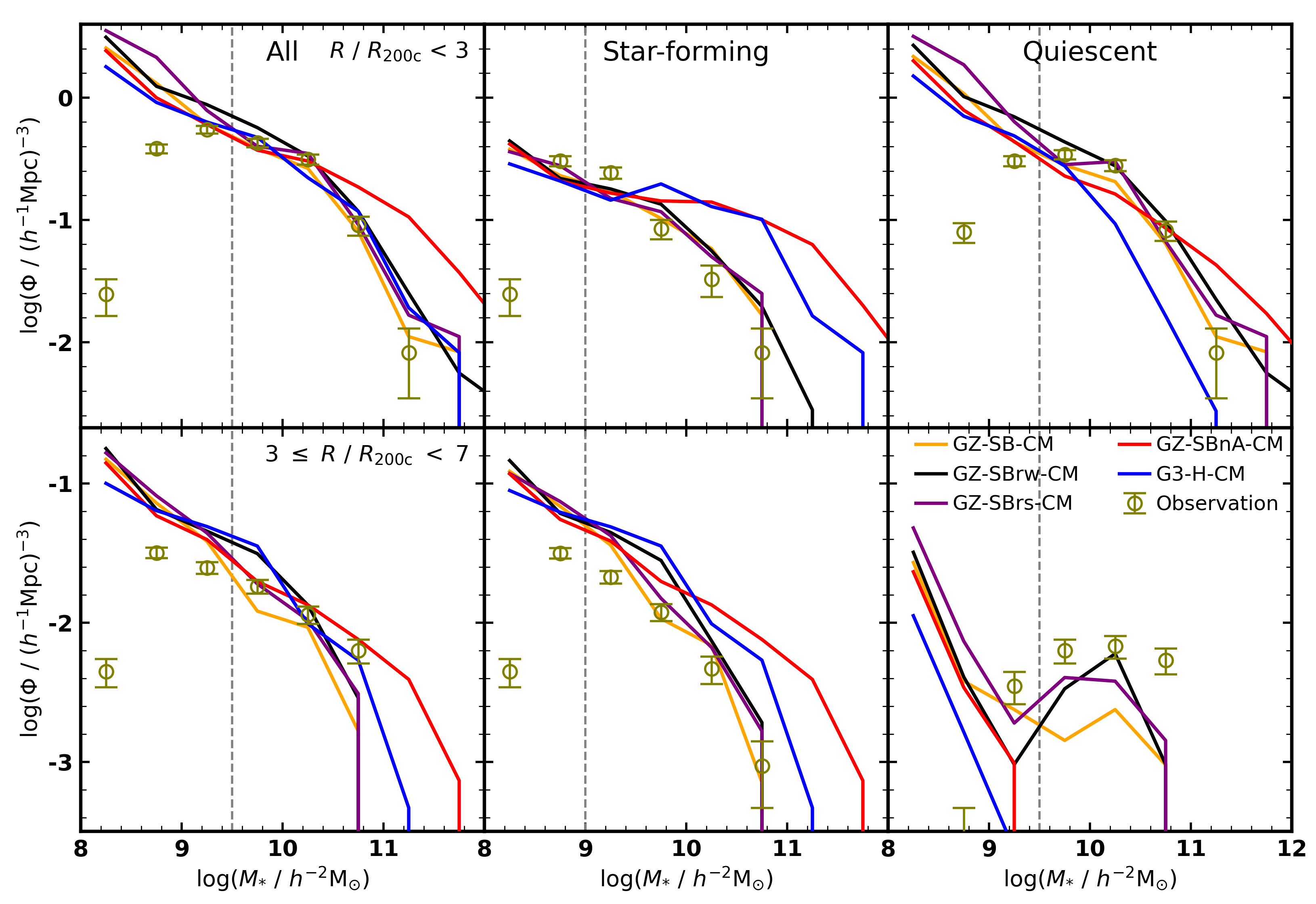}
	\caption{Stellar mass functions for galaxies in the high-density region (HDR; $R/R_{200c}<3$, upper panels) 
 and the low-density region (LDR; $3 \leq R/R_{200c}<7$, lower panels). The first column shows results for all galaxies, 
 while the second and third columns show results for star-forming and quiescent galaxies, respectively. 
 Different lines show results for different simulations as indicated in the lower-right panel, while 
 observational results are shown by symbols with Possion error bars. The vertical gray dashed lines represent 
 the estimated complete mass limit of the observations. } \label{fig_SMF_diffR}
\end{figure*}

\begin{figure}
   \centering
   \includegraphics[scale=0.36]{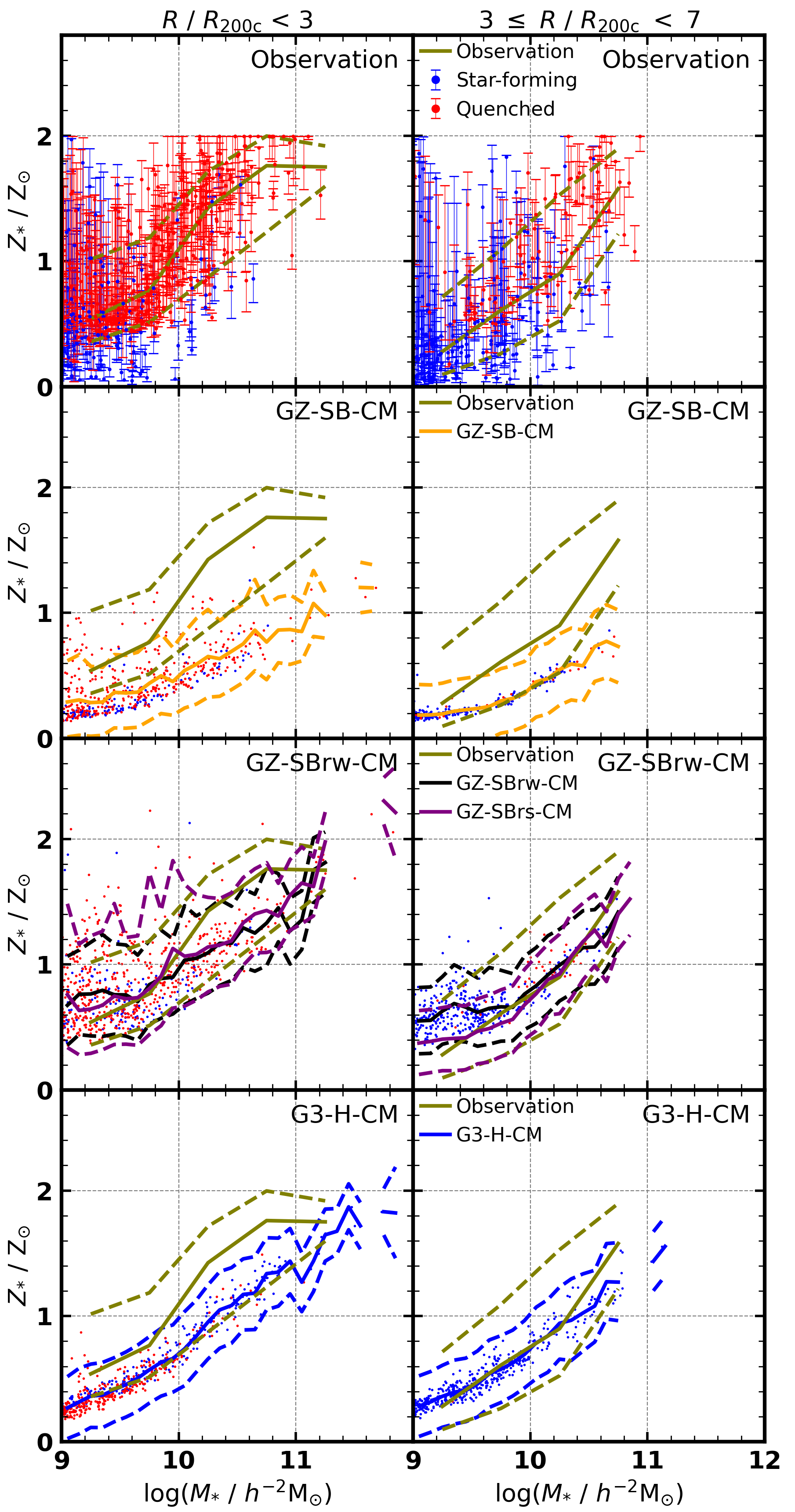}
	\caption{The $M_{*}$ - $Z_{*}$ relation of galaxies in HDR (left panels) and in LDR (right panels). 
 Red and blue dots represent quiescent and star-forming galaxies respectively.  The first row shows observational data, 
 and the rest rows show the results of simulations as indicated at upper right corner of each panel. The solid lines are the median relations, while 
 the dashed lines show the $16^{\rm th}$ - $84^{\rm th}$ percentiles of the distribution.  The scatter in the simulation relations represents 
 measurement uncertainties (see the text for details).  For comparison, the observational median relation and scatter 
 (brown color) are repeated in the other panels. In the third row, we also show the median relation and scatter 
 predicted by \GZSBrs -CM.}\label{fig_M_Zstar}
\end{figure}

\begin{figure}
   \centering
   \includegraphics[scale=0.36]{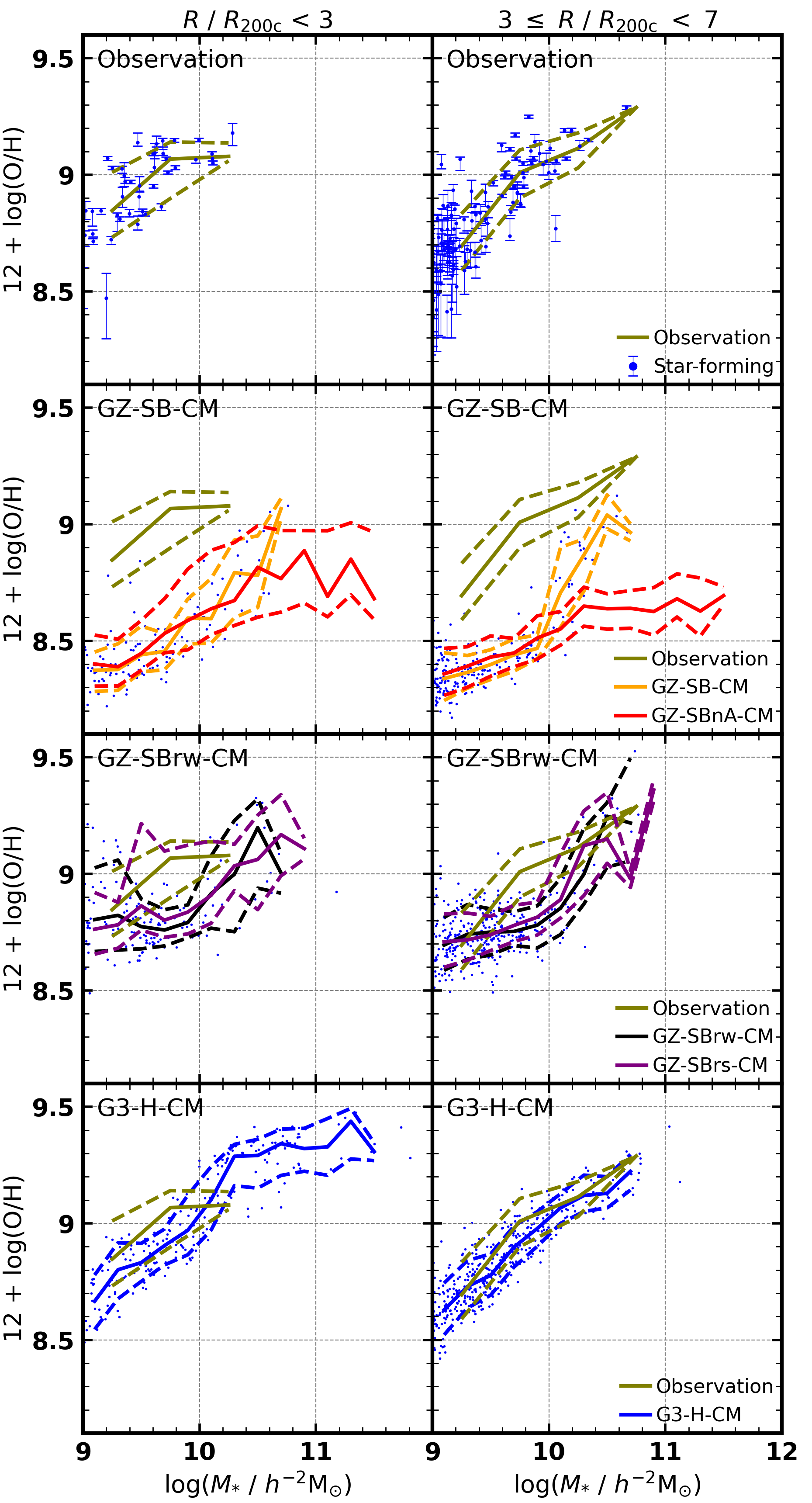}
	\caption{Similar to Figure \ref{fig_M_Zstar}, but for the $M_{*}$ - $12 + {\rm log(O/H)}$ relation of star-forming galaxies, which is referred to as the $M_{*}$ - $Z_{\rm ISM}$ relation in this paper}.\label{fig_M_Zgas}
\end{figure}

Figures \ref{fig_SMF_diffR}, \ref{fig_M_Zstar} and \ref{fig_M_Zgas} show the stellar mass function (SMF), 
stellar mass-stellar metallicity relation (\MSZR) and stellar mass-ISM metallicity relation (\MIZR) of 
observed galaxies in two different regions, HDR (within $3R_{\rm 200c}$ of the Coma cluster) and LDR
(between $3R_{\rm 200c}$ and $7R_{\rm 200c}$). 
In general, galaxy properties in the two regions differ significantly. Galaxies in HDR tend to be more massive, 
more frequently quenched, and metal richer than those in LDR, suggesting that environmental processes have a
significant impact on galaxy properties \citep{WangH2018, WangE2018}. It is thus interesting to examine whether or not the two fiducial simulations 
can reproduce these galaxy properties and their environmental dependencies. 

Let us first look at the SMFs.  Since all galaxies are located within a small volume, the SMFs can be obtained by directly 
counting the number of galaxies without any correction.  As shown in Figure \ref{fig_SMF_diffR}, the observed 
star-forming galaxies are complete to $10^{9.0}\msunt$ while quiescent galaxies are complete to $10^{9.5}\msunt$. So the total 
population is complete to $10^{9.5}\msunt$.
Note that the stellar mass of simulated galaxies with 200 star particles roughly corresponds to $\log(M_*/\msunt)=8.7$, lower than the observational limit.
In HDR, the predicted SMFs of the two simulations for the total 
galaxy population agree well with the observational data at $\log(M_*/\msunt)>9.5$. 
In LDR, the G3-H model reproduces the total SMF at $\log(M_*/\msunt)>9.5$,  while the GZ-SB-CM SMF is lower than the observation.  

Inspecting the quiescent and star-forming galaxies separately, we can see a larger difference between
observational and simulation results. The G3-H model produces more star-forming galaxies and fewer quiescent 
galaxies in both HDR and LDR than in the observation. This could be because the G3-H model does not include AGN feedback, 
but includes strong stellar and supernova feedback to prevent the growth of massive galaxies. 
Interestingly, the discrepancy, in the massive galaxies ($\gtrsim 10^{10} \msunt$), for star-forming galaxies is more prominent in HDR than in LDR, while the 
discrepancy for quiescent galaxies is the opposite. In particular, there are almost no quiescent galaxies in LDR, 
which is very different from the observation. This indicates that most of the quiescent galaxies in 
the G3-H-CM should be produced primarily by environmental effects, such as ram pressure and tidal stripping, while 
environmental effects are too weak to quench these massive galaxies in low-density regions. This large discrepancy seen 
in LDR suggests that additional quenching mechanisms, such as AGN feedback, are probably necessary to quench massive galaxies. Moreover, the ``over'' quenching of low-mass galaxies ($\lesssim 10^{9} \msunt$) may owe to the observational and simulation limits.

The GZ-SB model produces results that are much closer to the observational data than the G3-H model. 
In HDR, the predicted SMFs for both star-forming and quiescent galaxies match the observational results well. 
In LDR, the GZ-SB model works well for star-forming galaxies, but significantly underestimates the number of 
quiescent galaxies, with a difference of about 0.6 dex. This discrepancy is only visible when galaxies are 
separated according to their environments, as the number of quiescent galaxies in LDR is much lower 
than that found in HDR. This suggests that modifications of the subgrid physics are needed to reproduce
the observed properties of the Coma cluster and that constrained simulations 
can provide important constraints on galaxy formation models, as we will demonstrate below using more
observations.

Next, we examine the stellar mass-metallicity relation, shown in Figure \ref{fig_M_Zstar}.
Metallicity measurements of observed galaxies have quite a large uncertainty, as shown in the top row. 
Hence to compare the observation with our simulation results,
we assign the same observed metallicity uncertainties to our simulated galaxies. Specifically, for each simulated galaxy of a given mass, 
we generate 100,000 mock galaxies with their metallicity following a Gaussian distribution. This distribution has the same mean metallicity as the simulated galaxy
and a dispersion equal to the mean uncertainty of observed galaxies with the
same stellar mass as the simulated galaxy. The solid line 
shows the median metallicity of our mock galaxies as a function of stellar mass and the dashed 
lines indicate the $16^{\rm th}$ - $84^{\rm th}$ percentiles of the distribution. Results for the observed galaxies are 
repeated in panels of the simulated results for comparison.

The median \MSZR\ for the G3-H model in LDR is in good agreement with the observational data, 
and the predicted scatter is only slightly smaller than observed. However, the predicted \MSZR\ in HDR 
is similar to that in LDR and is systematically lower than the observational data. 
This appears in conflict with the SMF result, which shows a strong environmental dependence 
(Figure \ref{fig_SMF_diffR}). 
In addition, quiescent galaxies and star-forming galaxies in HDR follow the same trend, 
which is not consistent with observations \citep{Peng2015}. One potential explanation is that 
the strong stellar wind feedback in the G3-H model makes both the ISM and CGM of simulated 
galaxies more extended and hence more susceptible to ram pressure stripping, 
leading to a rapid removal of the ISM and little evolution in the stellar metallicity. 
This effect is more pronounced for low-mass galaxies and could owe to their shallower
gravitational potential wells, which would give a higher quenching rate for 
low-mass galaxies than for massive galaxies in HDR (see Figure \ref{fig_SMF_diffR}). 

The GZ-SB model predicts \MSZR s that are significantly different from the observational data, particularly at the massive end. 
This discrepancy in the same mass range is also visible in the original \simba\ paper \citep[figure 10 in][]{dave2019simba}. 
In LDR, the intrinsic scatter of the simulated relation is very small, similar to that of the G3-H model. In HDR, however, the 
intrinsic scatter is much larger, which is different from the G3-H model. This could 
owe to the fact that in a high-density environment, the CGM of a galaxy is more easily stripped while its ISM is not,
allowing the stellar metallicity to increase gradually over time owing to metals produced by supernovae. 
This suggests that the stripping effect in the low-mass GZ-SB galaxies is weaker than that in low-mass G3-H galaxies.

Figure \ref{fig_M_Zgas} displays the oxygen abundance of the ISM as a function of stellar mass for both HDR and LDR. 
We only present results for star-forming galaxies, as the ISM metallicity of real galaxies is measured using emission 
lines. The observed slope is steeper for low-mass galaxies than for massive galaxies, which is consistent 
with results found by \cite{Tremonti2004}. In addition, the ISM metallicity for low-mass galaxies is slightly 
higher in HDR than in LDR. The ISM metallicity of simulated galaxies is estimated using star-forming gas particles 
with a non-zero star formation rate, which might not be exactly the same as the observational definition.
The solid lines indicate median relations, and the dashed lines represent the 
scatter ($16^{\rm th}$ - $84^{\rm th}$ percentiles) of the relation.  Note that the scatter of the simulated \MIZR\ also involves
measurement uncertainties of the ISM metallicity, similar to the \MSZR\ described above. 

The \MIZR\ for G3-H galaxies in LDR is in agreement with the observational data in terms of both the overall trend 
and the scatter. In particular, the G3-H model predicts a flatter relation at the massive end, consistent 
with the observations. In HDR, the predicted metallicity is slightly lower than that observed for star-forming galaxies 
with masses $9<\log (M_*/\msunt)<10$, likely caused by the strong stripping effect discussed above. 
The simulated relation extends to more massive galaxies than the observed one, as the simulation predicts many 
massive star-forming galaxies at $\log (M_*/\msunt)>10$. Similarly to the \MSZR , the predicted \MIZR\ does not show any 
significant dependence on the environment.  Unlike the G3-H model, the GZ-SB predicts an ISM with metallicity 
that is significantly lower than the observed in both LDR and HDR, with the deficit being particularly large 
for low-mass galaxies (about 0.5 dex). This deficit likely has the same origin as the deficit in the stellar 
metallicity. Additionally, the predicted \MIZR\ is steeper at the high mass end than at the low mass end,
contrary to observational results.

\subsection{Intracluster and Intergalactic media}\label{sec_ICMIGM}

\begin{figure*}
   \centering
   \includegraphics[scale=0.5]{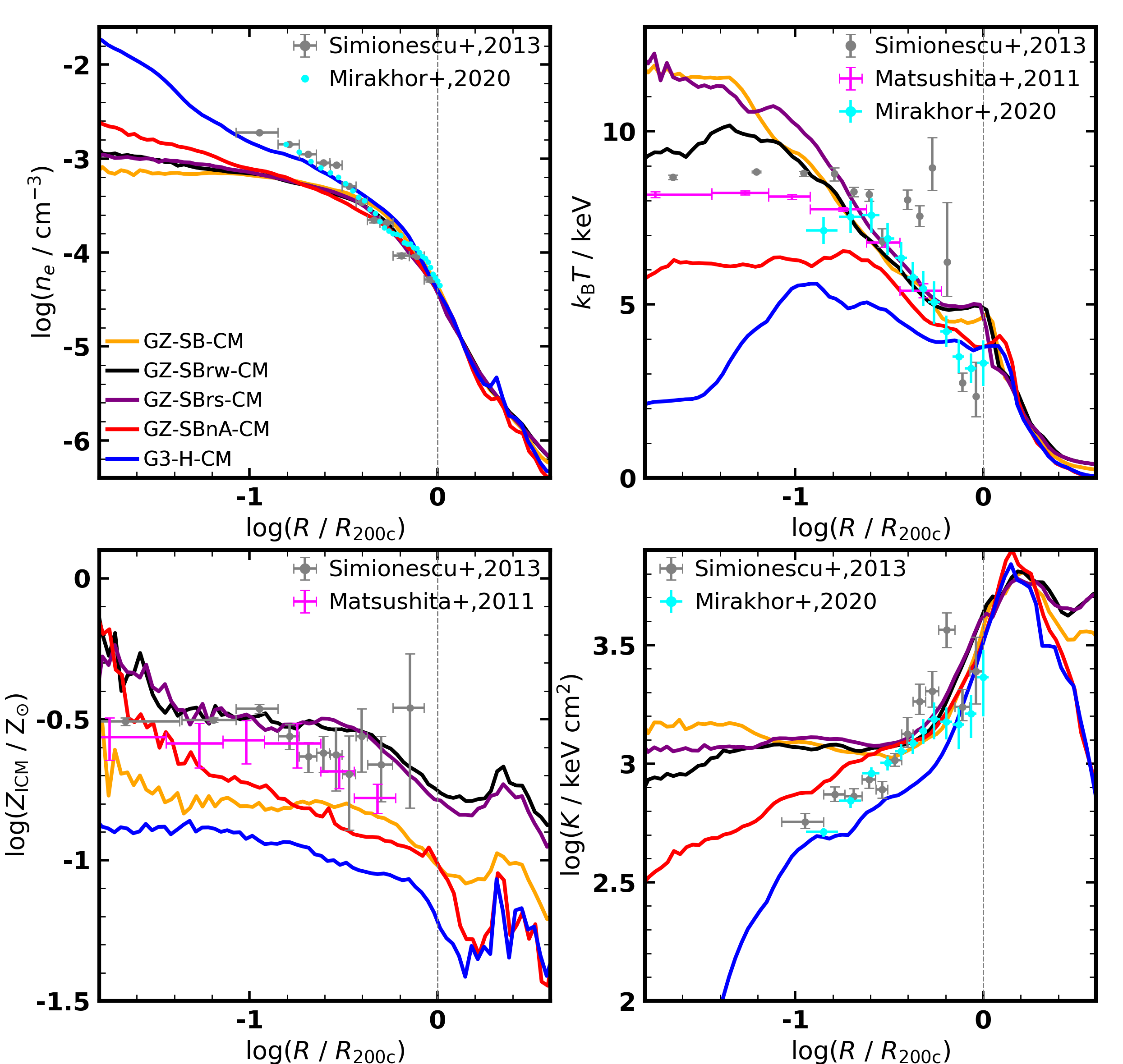}
	\caption{ICM profiles for the Coma cluster.  The four panels show electron number density (top left), 
 temperature (top right), metallicity (bottom left) and entropy (bottom right) profiles.  Symbols show 
 observational results taken from the literature as indicated in each panel, while lines plot the simulations results. 
 See Section \ref{sec_obs} for details of the observational data.}\label{fig_ICM_profile}
\end{figure*}

\begin{figure*}
   \centering
   \includegraphics[scale=0.55]{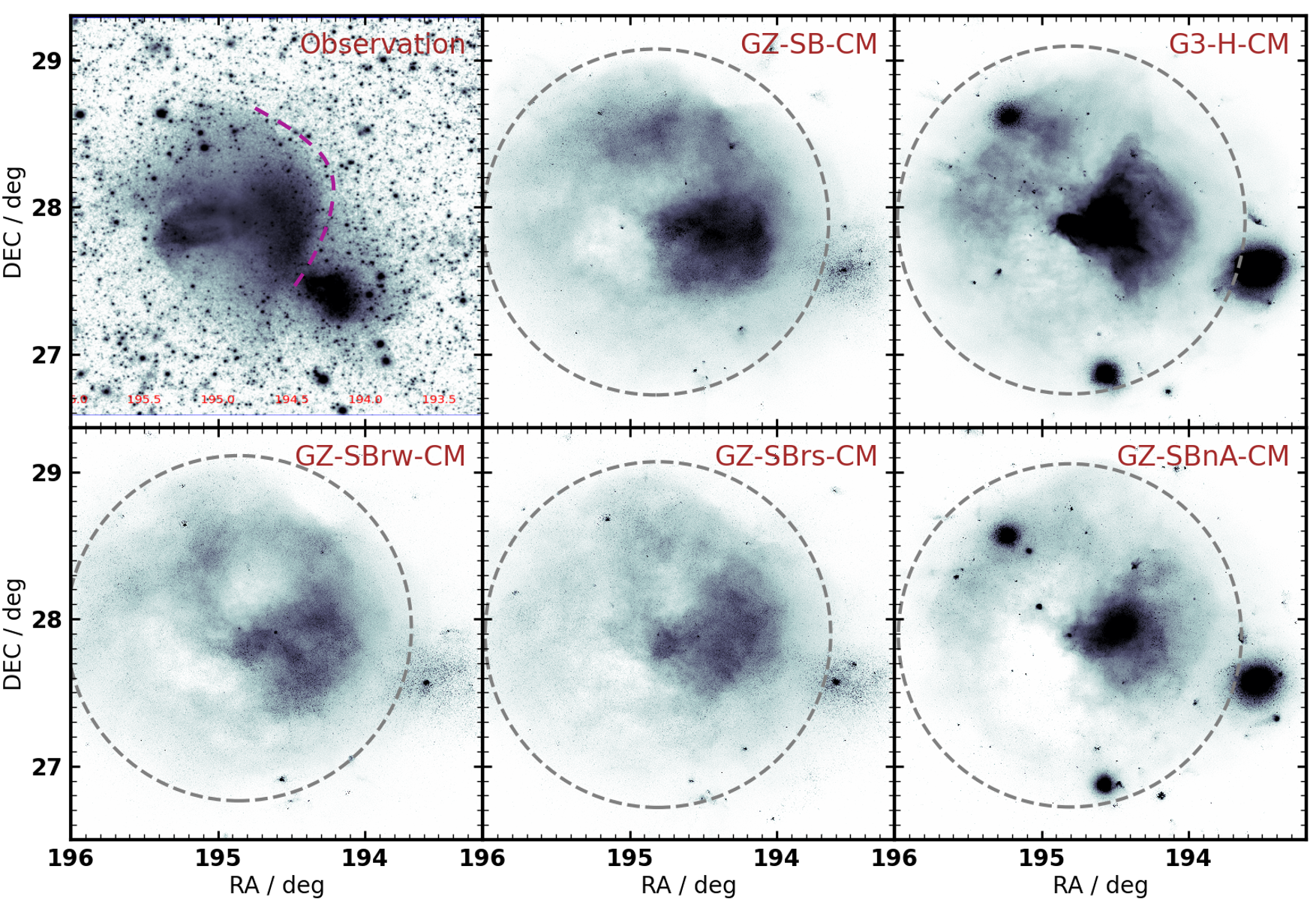}
   \caption{Scaled X-ray maps in the observations and the simulations as indicated (see the text for details). The observational data 
 is taken from \citet{Churazov2021A&A}. The purple dashed curve marks the observational shock front with a small shift. The southwest structure corresponds to the NGC 4839 subcluster. Note that there is a small offset between the simulated and real subclusters.  The simulated maps adopt the same color scale as the observational map (private communication for the observational scale). The gray dashed circles represent the scale of $R_{\rm 200c}$ in each simulation.} \label{fig_xraymap}
\end{figure*}

The intracluster medium (ICM) and intergalactic medium (IGM) can fuel star formation in galaxies, and the feedback 
from galaxies can also leave imprints on these media.
Therefore, it is essential to compare simulation predictions 
and observations of the gas they represent. Figure \ref{fig_ICM_profile} shows the observed 
electron density ($n_{\rm e}$), electron temperature, metallicity and entropy profiles of the ICM of the Coma 
cluster, taken from \cite{matsushita2011radial, Simionescu2013ApJ, Mirakhor2020MNRAS}. The electron density 
decreases rapidly with increasing distance from the cluster center, while the temperature decreases slowly, 
with a typical temperature of about 5 keV within the virial radius. The ICM entropy increases rapidly towards 
the cluster outskirts, and the ICM metallicity is roughly constant with radius, around $0.3\Zsun$. 
Note that different observations give quite similar results, suggesting that the observational data are reliable.

The simulated profiles are depicted as lines in Figure \ref{fig_ICM_profile}. The values are the average in 
3-dimensional spherical shells. As discussed in Section \ref{sec_obs}, the situation in observational results is 
more complex. The electron density and entropy profiles in Figure \ref{fig_ICM_profile} are deprojected, allowing direct 
comparisons with the simulation results. The temperature profile from \cite{Mirakhor2020MNRAS} is also deprojected, 
while others are not. Despite this, the profiles are similar, indicating that the projection effect are 
not significant.  Note that the temperature profile of \cite{Simionescu2013ApJ} is measured in a narrow sky 
region, which may explain the large fluctuation seen in the data. The observed metallicity profiles are not 
deprojected. Since the metallicity gradient is small, the projection effect may be ignored.

As one can see, the electron density profile predicted by the G3-H-CM simulation roughly matches 
the observational data, although it is slightly lower (higher) than the observation at 
$R<(>)0.3R_{\rm 200c}$. The GZ-SB model produces a density profile that is similar to the observational 
data at $R>0.3R_{\rm 200c}$, but has a much lower and flatter profile at $R<0.3R_{\rm 200c}$. 
Recently, \cite{robson2020x} found that massive clusters in the \simba\ flagship simulation show flat 
central gas cores similar to what we find here. This implies that flat cores are very common in \simba\ clusters, which could be due to its strong AGN feedback \citep[see a similar result in][ with the300 cluster sample with their hypothesis]{LiR2023}. 
The situation in the temperature profile is different. The GZ-SB-CM simulation produces a temperature profile 
that is in agreement with the observational data from 9 keV at $0.1R_{\rm 200c}$ to 3 keV at the virial radius. 
The only difference appears in the innermost region ($<0.1R_{\rm 200c}$), where the simulated temperature 
reaches 12 keV, higher than the observed value. The G3-H-CM simulation, on the other hand, predicts a much lower 
temperature within the half of the virial radius, with the difference becoming smaller near the halo boundary. 
This is likely because the cooling of the ICM is not effectively suppressed in G3-H-CM.

Employing the same observational method of \cite{Simionescu2013ApJ, Mirakhor2020MNRAS}, we calculate the ICM entropy using 
the definition $K(R)= P_{\rm e}(R)/n_{\rm e}(R)^{5/3}$, where $P_{\rm e}(R)$ and $n_{\rm e}(R)$ are the electron 
pressure and number density profiles, respectively. We assume an adiabatic gas and adopt $\gamma=5/3$, so that
$P_{\rm e}(R)=(\gamma-1)u_{\rm e}(R)$, where $u_{\rm e}(R)$ is the mean internal energy density of electrons at $R$. 
Both simulations generate steep entropy profiles in the cluster outskirts ($R>0.4R_{\rm 200c}$), in agreement with both 
the observations and previous results \citep[][]{pratt2010gas}. In the inner region ($0.1R_{\rm 200c}<R<0.4R_{\rm 200c}$), 
the GZ-SB-CM shows a much higher and more flat entropy profile than the G3-H-CM. The observational data lies between 
the two simulated profiles and appears to be closer to G3-H-CM. The discrepancy between the observation and 
GZ-SB-CM increases with decreasing radius. In the innermost region ($R<0.1R_{\rm 200c}$), the entropy profile
predicted by G3-H-CM drops quickly, while that predicted by GZ-SB-CM remains constant. This is likely owing 
to the presence or absence of AGN feedback in the two simulations.

As shown in Figures \ref{fig_M_Zstar} and \ref{fig_M_Zgas}, the stellar and interstellar metallicities predicted by 
the GZ-SB model are lower than the observed values. Where are the missing metals? Are they expelled into the ICM/IGM by 
feedback processes or do the galaxy formation models produce too few metals? The ICM metallicity is crucial for 
distinguishing between these two possibilities. As shown in Figure \ref{fig_ICM_profile}, the GZ-SB-CM produces a 
nearly flat metallicity profile within the virial radius and the average metallicity is around $0.16\Zsun$, which is 
about half of the observed value. The discrepancy between the observation and the G3-H-CM simulation is even greater, 
with the predicted mean metallicity only about $0.13\Zsun$. These results demonstrate that the star formation 
processes used in these models produce too few metals. In the next section, we will improve the model predictions 
with new simulations.

\cite{Li2022ELUCID} found a bow-like shock in the G3-H-CM Coma cluster, which was also observed in the same 
location by eROSITA \citep{Churazov2021A&A}. Figure \ref{fig_xraymap} displays the shock feature in the 0.4$-$2 keV band 
observed by eROSITA, taken from Figure 6 in \cite{Churazov2021A&A}. To compare the simulations with the observation, 
we show X-ray brightness maps for the simulated Coma in the same band. The method for calculating the X-ray maps 
is described in \cite{Li2022ELUCID}. To emphasize the shock feature, \cite{Churazov2021A&A} applied a flat-fielding 
procedure by normalizing the original image with a mean model profile. To mimic the observation, a background 
X-ray emission was added and a similar procedure was adopted in the simulated maps. It is evident that the 
bow-like shock front in G3-H-CM is very prominent, while the feature is weaker in GZ-SB-CM. The formation of a 
shock front is sensitive to gas properties before the collision of gas flows \citep{Zinger2018MNRAS}, 
thus providing an opportunity to explore the ICM properties. In addition, the substructure seen in the 
southwest of the Coma cluster in the eROSITA observations is also present in the X-ray map of the G3-H-CM, 
but almost absent in that of the GZ-SB-CM.

\begin{figure*}
   \centering
   \includegraphics[scale=0.8]{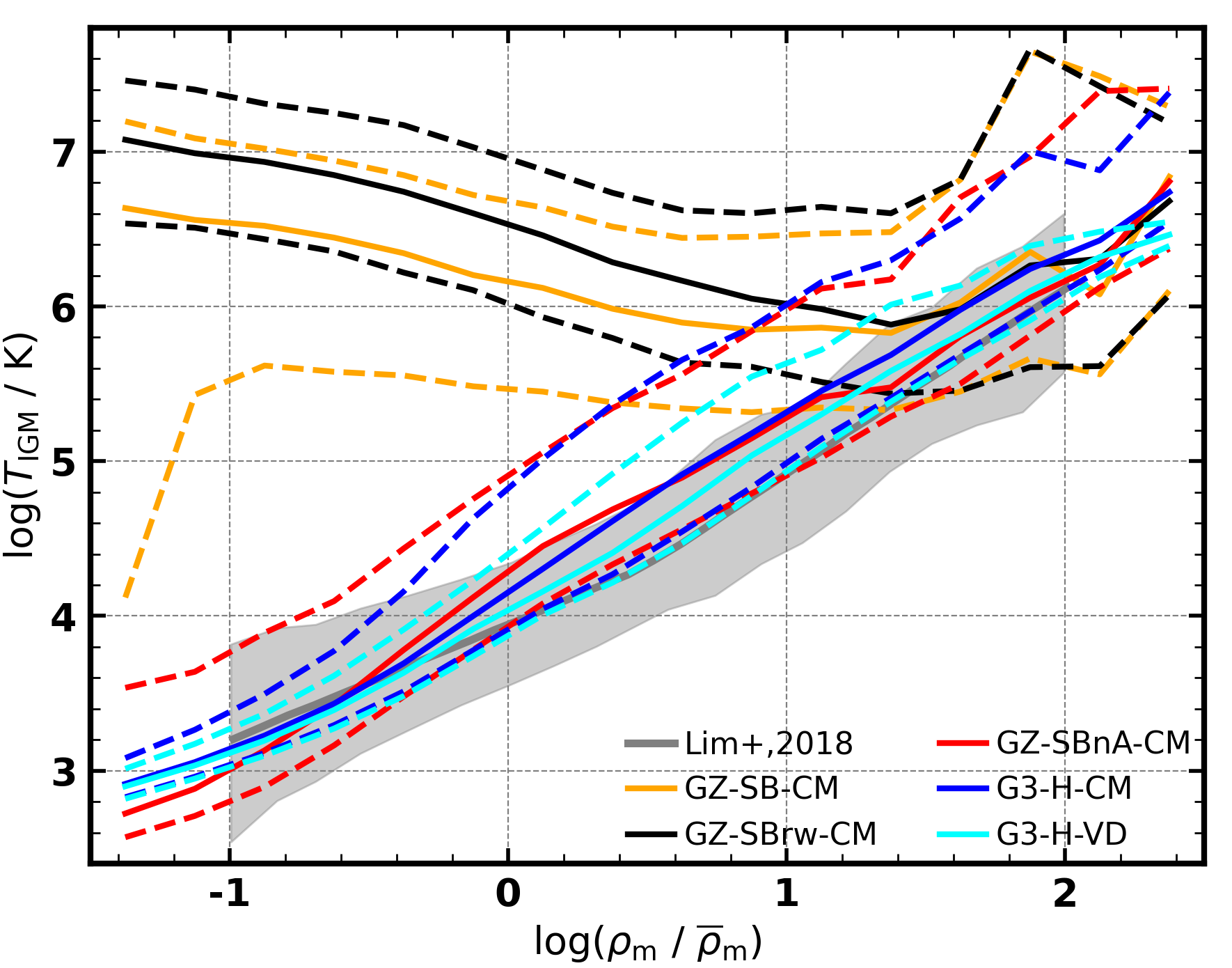}
\caption{The $T_{\rm IGM}$ - $\rho_{\rm m}$ relation.  The gray line shows the observational relation from \cite{Lim2018ApJ}, 
with the shaded region indicating its uncertainty. The other solid lines show the median relations obtained from the different simulations as indicated in the figure, 
and the dashed lines correspond to the $16^{\rm th}$ - $84^{\rm th}$ percentiles of the distributions.}\label{fig_Trhorelation}
\end{figure*}
 
The lower-right panel of Figure \ref{fig_ICM_profile} shows that the entropy at $R>4R_{\rm 200c}$ is significantly 
different between the two fiducial simulations. This implies that the IGM property may also be used to distinguish 
different models. Combining the reconstructed density field \citep{WangH2009MNRAS, WangH2016ApJ} and the {\it Planck} SZ map, 
\cite{Lim2018MNRAS} obtained the IGM temperature ($T_{\rm IGM}$) as a function of mass density ($\rho_{\rm m}$), 
where the temperature and density are both estimated on grids of a size of $1\mpc$. 
The result is shown in Figure \ref{fig_Trhorelation}, with the shaded region representing the 
uncertainty of the relation. The observational $T_{\rm IGM}$ increases with increasing $\rho_{\rm m}$ 
from $\log (T_{\rm IGM}/{\rm K})\sim3$ at  $0.1\bar\rho_{\rm m}$ to $\log (T_{\rm IGM}/{\rm K})\sim6$ at $100\bar\rho_{\rm m}$. 
We note that the result of \citet{Lim2018MNRAS} was derived using data from the entire SDSS survey region, 
rather than just the Coma region, and so the comparison with simulation results is with this caveat. 

To compare with the observation, we divide the HIR into grids of side length $1\mpc$. For each grid, 
we calculate $T_{\rm IGM}=\bar P_{\rm e}/k_{\rm B}\bar n_{\rm e}$, where $\bar P_{\rm e}$ and $\bar n_{\rm e}$ are the average 
electron pressure and electron density of the grid, respectively, and $k_{\rm B}$ is the Boltzmann constant. 
Here $\bar n_{\rm e}$ is calculated directly from $\rho_{\rm m}$ and the cosmic baryon fraction, assuming zero 
metallicity and full ionization \citep[see][Section 4.2]{Lim2018MNRAS}. This mimics the observation as the SZ 
$y$ parameter is the integral of electron pressure. In this analysis, we exclude contributions of 
star-forming gas particles and wind particles. As before, the simulation results in Figure \ref{fig_Trhorelation} 
are presented so that the median $T_{\rm IGM}$-$\rho_{\rm m}$ relations are shown 
by solid lines and their $16^{th}$ and $84^{th}$ percentiles by dashed lines. As one can see, the IGM temperature predicted by 
the GZ-SB-CM first decreases and then increases with the density. The GZ-SB-CM result is consistent with the 
observational data at $\rho_{\rm m}>10\bar\rho_{\rm m}$. At $\rho_{\rm m}<10\bar\rho_{\rm m}$, however, a deviation 
appears and increases with decreasing $\rho_{\rm m}$. At $\rho_{\rm m}=0.1\bar\rho_{\rm m}$, the GZ-SB-CM predicts 
a $T_{\rm IGM}$ that is about three orders of magnitude higher than the observations. In contrast, the predicted 
$T_{\rm IGM}$ increases monotonically with $\rho_{\rm m}$ in G3-H-CM, with both the amplitude and slope of the 
relation in good agreement with the observational data.

\section{Using the Coma Cluster as a Test Bed to Diagnose Feedback Processes} 
\label{sec_calimp}

In the preceding section, we see a complex situation in model predictions. Neither simulation can accurately reproduce all the observational results, and each of the simulations has its advantages and disadvantages. 
These simulations are tailored to fit observational data in a large volume, but almost no data on ICM and IGM 
were used to constrain model parameters. Our analyses above revealed two main discrepancies: the first is the 
under-prediction of metallicities in stars, the ISM and the ICM (the metal budget problem), and the second is the 
discrepancies in the physical properties of the ICM and IGM. In this section, we attempt to understand the underlying 
mechanisms that cause the discrepancies and to make improvements in the related modeling.  

\begin{figure*}
   \centering
   \includegraphics[scale=0.38]{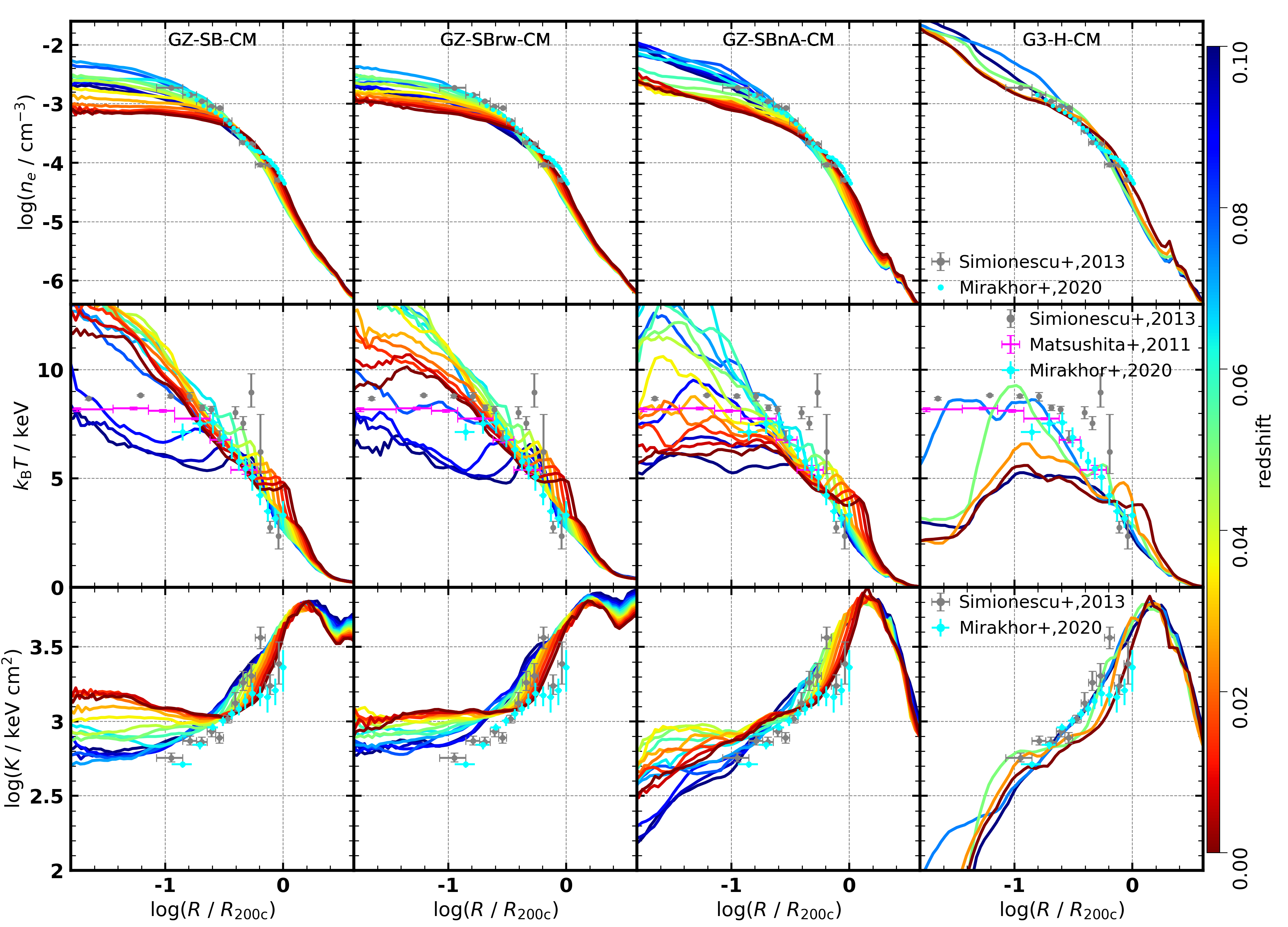}
	\caption{ Evolution of the Coma ICM profiles from $z=0.1$ (blue) to 0 (red) as indicated by the color-coded lines.  
 Different columns show results for different simulations, as indicated in the first row. The first row shows 
 electron density, the second temperature and the third entropy.  Data points are observational results taken from 
 the literature (the same as Figure \ref{fig_ICM_profile}). } \label{fig_ICM_profile_history}
\end{figure*}

\subsection{Metal budget problem and model calibration}\label{sec_calib}

We first examine the metal budget problem for the GZ-SB model. The metallicities of stars, ISM, and ICM can all 
be affected by a variety of processes. To determine the source of the metal budget problem, it is essential to 
distinguish the effects produced by different processes. As seen in Figures \ref{fig_M_Zstar} and \ref{fig_M_Zgas}, 
this problem is present in both HDR and LDR, and the discrepancy between the observations and the model is similar 
in both regions. This suggests that internal processes are responsible for the problem, since environmental effects 
are expected to be much weaker in LDR than in HDR. Furthermore, the problem is significant in both low-mass and 
high-mass galaxies, indicating that the relevant process should not depend strongly on stellar mass. This eliminates 
AGN activity and feedback as the cause, as AGN feedback is expected to be important only for massive galaxies. 
The predicted amounts of metals in all baryonic components, i.e. stars, ISM, and ICM, are all lower than the 
observed values, suggesting that we should increase the total metal yield rather than 
redistributing it among the different baryonic components.

Thus, one potential solution to the metal budget problem is to increase the star formation rate.  In the two 
modified models, \GZSBrw\ and \GZSBrs , we adopt a slightly higher star formation efficiency per free-fall time,  
${\rm SFR}/(M_{\rm gas}/t_{\rm freefall}) = 0.03$, instead of 0.02 used in the fiducial \simba\ model.  Additionally, 
a slightly higher metallicity floor is adopted to raise the H$_2$ fraction, to increase the SFR at 
high redshift to meet the recent JWST findings of very high redshift galaxies. To counterbalance the mass 
growth caused by the increased SFR, the supernova strength is also enhanced through the coupling factor 
and the wind efficiency, which should also increase the metals in the ISM and ICM. The under-prediction of 
quiescent galaxies in LDR suggests that the AGN feedback 
should also be increased, which is consistent with the results from The 300 Clusters Project 
\citep{Cui2018MNRAS,cui2022three}. To this end, we adopt a much higher jet velocity of 15000 km/s in the 
strong Jet model -- \GZSBrs , instead of 7000 km/s in the fiducial run. In the weak Jet model -- \GZSBrw , we employ 
a different strategy that changes the AGN feedback efficiency by shifting the jet onset time to an earlier redshift, 
which we accomplish by heating the wind particles to the virial temperature with a jet velocity similar to the fiducial model.  
\GZSBrw\ and \GZSBrs\ have the same star formation and supernova feedback model but different AGN feedback 
models (Table \ref{TableSimulations}). In addition, we include the no AGN run of the fiducial \simba\ model 
to examine the effect of AGN feedback.

Figure \ref{fig_M_Zstar} shows that our modification has a significant effect on the \MSZR . Compared to the fiducial model, 
the stellar metallicity of the entire galaxy population in \GZSBrw -CM increases. The increase is more pronounced 
for massive galaxies than for less massive ones, resulting in a steeper slope. The \MSZR s in both HDR and LDR are 
now in line with the observational data. The scatter of the simulation result, taking into account observational 
uncertainties, is slightly smaller than the observational one, indicating that the intrinsic scatter of real 
galaxies is only slightly larger than that of the simulated galaxies. The results for \GZSBrs -CM can be used to 
assess the impact of AGN feedback on the \MSZR . It is evident that the \MSZR\ in LDR is very similar to that of 
\GZSBrw -CM. In HDR, it produces a slightly larger scatter, suggesting that AGN feedback does not have a major 
effect on the result.

The results of the \GZSBrw , \GZSBrs\ and G3-H models indicate that the intrinsic scatter of the \MSZR\ in low-density regions 
is small. This suggests that the large scatter observed in the data mainly owes to measurement uncertainties. 
In higher-density regions, however, the intrinsic scatter is larger, as environmental processes can impede gas accretion 
and raise the ISM and stellar metallicity. This demonstrates that if galaxies in different environments are studied together, 
a tight correlation will not be observed. Therefore, the study based on constrained simulations of the same 
Coma cluster provides a unique way to distinguish different mechanisms that contribute to the \MSZR . 
In a future study, we will investigate how the \MSZR s in different regions are formed and how the ICM of the Coma cluster 
is enriched based on these simulations.

The revision also improves the prediction of the ISM metallicity, as seen in Figure \ref{fig_M_Zgas}. 
The predicted oxygen abundance increases by approximately 0.4 dex at $\log (M_*/\msunt)=9$ and by 0.2 dex 
at $\log (M_*/\msunt)=10.5$. The two new simulations yield \MIZR s that are much closer to the observed values than 
the fiducial model. However, the difference is still significant. For example, there is a clear break in the 
predicted \MIZR\ at $\log (M_*/\msunt)\sim10$ and the slope of the \MIZR\ at the low-mass end is smaller than that at 
the high-mass end. To understand the origin of this break, we show results for GZ-SBnA-CM, which has 
disabled AGN activity (Table \ref{TableSimulations}). We can see that the GZ-SBnA-CM relation is very flat and 
consistent with the result for low-mass GZ-SB-CM galaxies. 
This demonstrates that the break in GZ-SB-CM, \GZSBrw -CM and \GZSBrs -CM, is caused by AGN feedback. 
In the observed star-forming galaxies, the slope decreases as $M_*$ increases, which is the opposite of what 
is seen in the simulations. Recently, \cite{Hong2023ApJ} found that quenching processes usually occur in 
dynamically hot systems. These star-forming galaxies are mainly dynamically cold, disk-like galaxies,
and so AGN feedback in them should be inefficient. This might explain why we do not see a similar signal 
in real galaxies. A more detailed study on this issue is needed to further our understanding of AGN feedback.

The discrepancy between the ICM metallicity in the original models and the observational data is eliminated 
in the two modified models. As seen in Figure \ref{fig_ICM_profile}, the two revised models produce flat ICM metallicity 
profiles around $0.3\Zsun$, which is consistent with the observational data in terms of both amplitude and 
radial dependence. Increasing the star formation rate and supernova feedback is necessary to 
match the observational results. The new models also predict a much higher metallicity of the IGM at $R>R_{\rm 200c}$ than 
the fiducial model, which could be tested through quasar absorption lines. Figure \ref{fig_2Dmap} shows the evolution 
of the ICM metallicity of the \GZSBrw -CM simulation. Little change is seen in the mean metallicity since $z=0.2$ when 
the Coma cluster was being formed. This might explain the flat slope of the ICM metallicity, as the metals had already 
been released to the ICM before the Coma cluster was formed.

We also examine the impact of the modification on the SMF. The SMF discrepancies between the GZ-SB-CM and the 
observations mainly appear in LDR. In the new models, the number density of quiescent galaxies is significantly 
enhanced in LDR.  Therefore, the deficit of quiescent galaxies becomes much smaller. In contrast, the number 
density in HDR is almost unchanged.  Comparing the two new models, we see that enhancing the jet velocity 
indeed increases the quiescent population, but the effect is small.
We compare the mass of star-forming gas ($M_{\rm SF}$) in star-forming galaxies before and after the 
modification. In LDR, star-forming galaxies in the \GZSBrw /rs model have lower $M_{\rm SF}/M_*$ than those 
in the GZ-SB model. The higher star formation rate and supernova feedback may consume and expel more cold 
gas than in the fiducial model. This may make these galaxies more vulnerable to AGN feedback, so as to 
increase the number of quenched galaxies with $\log (M_*/\msunt)>10$. In HDR, environmental effects may play the dominant 
role in quenching galaxies, so the increase of quiescent galaxies by the change in star formation rate and 
supernova feedback is insignificant.

We also examine the physical properties of the ICM and IGM predicted by the new models. As seen in 
Figure \ref{fig_ICM_profile}, the electron density and entropy profiles of the two new models are 
similar to those of the original GZ-SB model. The \GZSBrw\ model appears to be a better match to the 
observed temperature profile in the inner region ($R<0.1R_{\rm 200c}$) than both the GZ-SB and \GZSBrs\ models, 
likely because of its weaker AGN feedback. The bow-like shock in the two simulations is still very 
weak, as shown in Figure \ref{fig_xraymap}. In addition, the IGM temperature at low-density environments 
is much higher than the observed one, similar to the GZ-SB model (Figure \ref{fig_Trhorelation}). In general, 
the predicted ICM and IGM properties, except for gas metallicity, are comparable with the original simulations. 
Thus, the modifications made do not seem to resolve the discrepancies found in the ICM and IGM
(see the next subsection for further discussions).

Finally, we consider the metal budget problem for the G3-H model. The predicted \MSZR\ and \MIZR\ are in good 
agreement with the observations in LDR, but the discrepancy in HDR is significant smaller than that 
for the GZ-SB model. This is distinct from the GZ-SB model problems, and suggests that the modification strategy 
for the GZ-SB model would not work for the G3-H model. We believe that the issue is related to environmental effects. 
The similarity in the \MSZR\ between star-forming and quiescent galaxies suggests that 
environmental processes can quickly strip the ISM from satellites and prevent further star formation and 
metal production. A similar argument may also apply to the metal deficit in the ICM. One potential cause of 
the strong environmental effect is that strong stellar wind feedback makes the ISM vulnerable to tidal/ram pressure stripping 
\citep{Bahe2015MNRAS}. However, decreasing the feedback strength is not a solution as it will create too many 
massive galaxies. An additional internal quenching mechanism seems to be necessary.

\subsection{Sources of the discrepancy between simulations and observations in ICM and IGM properties}

There are a number of reasons why the simulated ICM/IGM properties are different from those observed, 
as suggested by Figures \ref{fig_ICM_profile}, \ref{fig_xraymap} and \ref{fig_Trhorelation}. 
The first is that the reconstruction may not be precise enough to recover substructures that 
depend sensitively on nonlinear processes. The second is that hydrodynamic solvers may not be able to 
handle complex and violent processes accurately. The third is that the subgrid physics, such as star formation, supernova wind feedback, and black hole accretion and AGN feedback, may not be appropriately modeled in the simulations, owing either to a lack of resolution, failures in the hydrodynamic scheme or inadequacies of the models or any combination of the above. Moreover, some mechanisms that are not included here, such as cosmic rays and magnetic fields, 
may also affect the ICM/IGM.
To better understand the effect of AGN feedback, we involve the GZ-SBnA-CM simulation (Table \ref{TableSimulations}), 
which turns off black hole accretion and AGN feedback. As one will see, constrained simulations based on
ELUCID provide an unprecedented opportunity to assess all these effects by comparing
predictions of different models with the observational data.

We first focus on the ICM within the Coma cluster. Figure \ref{fig_ICM_profile_history} illustrates the 
development of the density, temperature, and entropy profiles at $z<0.1$ for the GZ-SB, \GZSBrw , GZ-SBnA 
and G3-H models. This shows how these profiles today are established in the recent past.
A similar evolution history can be observed in all four simulations. At $z\sim0.1$, the temperature 
profile is nearly flat. The temperature in the inner region then increases rapidly, producing a 
declining temperature profile at $z=0$. There are only two major heating mechanisms for the ICM in the 
simulations: one is the shock heating caused by accretion and mergers, and the other is heating by AGN feedback. 
Since the two simulations without AGN feedback give comparable results, only shock heating seems to 
be able to drive the evolution in the temperature profile. As can be seen in both the dark matter and star 
maps shown in Figure \ref{fig_2Dmap}, there is indeed a massive substructure that is moving towards the Coma center 
at $z \sim 0.1$. 

It is interesting to examine how the ICM reacts to the merger event in the different simulations. 
Comparing GZ-SBnA-CM with GZ-SB-CM and \GZSBrw -CM, it is evident that GZ-SBnA-CM is similar to the other 
two in the outer region of the cluster ($0.4R_{\rm 200c}<R<1.5R_{\rm 200c}$) but significantly different in 
the inner region. Therefore, in the following analysis, we will focus on the inner region. At $z\sim0.1$, GZ-SBnA-CM has a much 
higher density, a slightly lower temperature, and a much lower entropy than the other two simulations. 
These differences likely owe to the presence or lack of AGN feedback. As the merger occurs, the density initially 
increases with time and then decreases, the temperature initially rises quickly with time and 
then remains more or less constant for the simulations with AGN feedback but quickly drops for the GZ-SBnA-CM, 
and the entropy increases quickly and forms a flat core in the central 
region in all three simulations. However, the density and entropy profiles in the inner region 
in GZ-SB-CM and \GZSBrw -CM are always flat, very different from those in GZ-SBnA-CM, in which the profile slopes 
vary significantly with time. This suggests that the AGN heating in \simba\ stabilizes the inner density 
at a relatively low level, and the entropy at a relatively high level. This is also seen in other state-of-art 
simulations \citep[][]{altamura2023eagle, lehle2023heart}.

Next, let us compare G3-H-CM with GZ-SBnA-CM. At $z\sim 0.1$, the G3-H-CM simulation has density, temperature, and entropy 
profiles that are similar to GZ-SBnA-CM: the density/entropy profiles are steep while the temperature 
profile is flat. After the merger, however, differences between the two become visible. The density and 
entropy of G3-H-CM remain more-or-less unchanged and stay close to the observed values. The temperature of 
G3-H-CM first increases quickly to reach the observed profile, and then quickly drops back to the value at 
$z\sim0.1$. In contrast, the temperature in GZ-SBnA-CM drops much more slowly. The different responses to the merger 
in G3-H-CM and GZ-SBnA-CM might be caused by the fact that \gadget\ and GIZMO codes handle hydrodynamics differently 
and that the cooling in \gadget\ is more efficient \citep[see also][]{Huang2019MNRAS}. 

The primary source of uncertainties in the reconstruction on halo scales is 
that highly nonlinear events, such as mergers, may not be well reproduced. 
In principle, one might investigate the effects of such uncertainties by simulating 
different realizations sampling the posterior distribution of reconstruction 
\citep[e.g.][]{Ata2022NatAs}. However, running many hydrodynamic simulations, especially at the resolution in this study, to achieve 
this is very costly. The basic property of ELUCID \citep{WangH2014ApJ,WangH2016ApJ} 
is that constrained simulations based on the reconstructed initial 
conditions can reproduce the large-scale structure at $z\sim 0$. Our N-body simulations 
showed that the large-scale structure around, and the mass of the Coma cluster   
have evolved little since $z\sim 0.1$, suggesting that each snapshot at $z<0.1$ 
may be taken to represent those in the present-day universe. The changes 
produced by merger events at $z\sim 0.1$ might then be considered as a rough 
measure of the fluctuations generated by reconstruction uncertainties. 

In general, the density and temperature profiles predicted by the four simulations 
are in line with the observational data within the `uncertainties' given by the variations 
among the snapshots at $z\sim 0.1$. In the inner region, the median density profiles of 
GZ-SB-CM and \GZSBrw -CM are lower than those observed, whereas the median profiles of 
the two non-AGN simulations match the observational data better. 
The predicted median temperature of the G3-H-CM is slightly lower than the 
observed value, while the predictions of the other three simulations match 
the observation. The most significant difference is seen in the entropy profiles.
The simulations that incorporate AGN feedback fail to reproduce the observed entropy 
profiles in the inner region, while the two simulations without AGN are
in good agreement with the observational data.
Despite the differences between \gizmo\ and \gadget, both G3-H-CM and GZ-SBnA-CM 
demonstrate that the CSs of the Coma cluster without including the AGN feedback can 
reproduce the observed ICM properties better than simulations that include the AGN feedback.
Interestingly, the density, temperature and entropy profiles at a slightly higher 
redshift, e.g. $z\sim0.05$ in GZ-SBnA-CM and G3-H-CM, match the observational results 
better than at $z=0$. Thus, if the reconstruction were fine-tuned so that the predicted 
merger occurred slightly later, the simulation results at $z=0$ would match 
the observational data better.

Figure \ref{fig_xraymap} shows the X-ray map around the Coma cluster and as discussed earlier, the map 
contains two prominent structures: a subcluster in the southwest with a small offset relative to the observed one and a shock front in the central region. 
The central galaxy of the subcluster is NGC 4839, with a stellar mass of $\log (M_*/\msunt)=11.07$ \citep[][]{Yang2007ApJ}. 
According to the stellar mass-halo mass relations of the empirical models \citep[e.g.][]{Behroozi2019}, 
the halo mass of the subcluster is estimated to be   
$\log (M_{\rm h}/\msun)=13.5$. The halo mass of the corresponding simulated subcluster is 
$\log (M_{\rm 200c}/\msun)=13.6$, consistent with the halo mass of NGC 4839. As shown in Figure \ref{fig_xraymap}, 
the X-ray emission from the subcluster is quite prominent in the observational data, 
and in the two simulations without AGN feedback, but is quite weak in the three simulations 
with AGNs. The X-ray luminosities of the NGC 4938 subcluster in G3-H-CM and GZ-SBnA-CM are comparable and 
about 10 to 20 times higher than those in the other three simulations with AGN feedback. 
\cite{Sasaki2016} estimated the gas mass and total mass within the central 98$\Kpc$ of the NGC 4839 subcluster 
using X-ray data and weak lensing, respectively. They obtained a gas fraction of 6.7\%.  The predicted gas 
fraction within 98$\Kpc$ is about 7.5\% for the G3-H-CM simulation and 3.7\% for the GZ-SBnA-CM simulation, 
but only about 0.5\% for the three simulations with AGN feedback. Apparently 
the AGN feedback implemented in our GZ-SB series is too effective in removing the CGM in 
the central region of the subcluster. 
Our analysis suggests that the CGM in group-sized halos can be 
significantly affected by AGN feedback, and that the X-ray emission of the CGM is a sensitive probe of this 
feedback. 

In the simulations, 
the inner shock feature seen in the X-ray maps is triggered by a collision 
between the main body of the Coma cluster and a subhalo. We found that the subhalo gas was largely 
expelled by the AGN feedback before the collision in GZ-SB-CM and \GZSBrw /rs-CM, so that the 
collision cannot generate a strong shock. The shock in GZ-SBnA-CM appears weaker than that in G3-H-CM, 
probably because a larger portion of the gas in the subhalo is converted into stars in GZ-SBnA-CM
(see SMF in Figure \ref{fig_SMF_diffR}) owing to a lack of AGN feedback in GZ-SBnA-CM 
and weaker supernova wind feedback than in the G3-H-CM simulation. 
The bow-like structure observed in the eROSITA X-ray data does not correspond to features
generated by an directional AGN jet. Typically, AGN jets create X-ray cavities by driving the 
ICM away \citep{Fabian2012ARAA}, while the observed structure is in fact brighter than the 
X-ray background produced by the ICM. The morphology of the structure closely resembles
the shock front generated by merger events in our constrained simulations \citep[e.g.][]{LiH2023}. 
This suggests that the amount of subhalo gas in the real universe should not have been 
reduced severely by feedback before the collision which generates the shock, 
because otherwise the collision would not be able to generate large amounts of X-ray. 
Unfortunately, a quantitative comparison with the observational data is not feasible due 
to the unavailability of the data. In reality, AGN activity is expected to 
occur sporadically and randomly, a behavior quite different from that expected 
from the roughly stable structure seen in the simulations.
A more stochastic AGN feedback model might result in a lower coupling with nearby gas elements, thereby exerting a lesser influence on the ICM.

Finally, we attempt to understand the source of the discrepancy in IGM properties (Figure \ref{fig_Trhorelation}).
We can see that \GZSBrw\ also predicts a much higher $T_{\rm IGM}$ than the observations at low density, similar to GZ-SB.
We do not present results of the \GZSBrs\ simulation, as the results for the two modified models are similar. This is expected, 
as the AGN feedback models in the GZ-SB model and its revisions are similar. We also display the result for GZ-SBnA-CM 
in Figure \ref{fig_Trhorelation}. As one can see, this result is very similar to that for G3-H-CM and 
the observational data. This suggests that the two codes, \gadget\ and GIZMO, can generate consistent IGM properties, 
as neither of them has AGN feedback. In addition, GZ-SBnA-CM and G3-H-CM implement very different stellar and supernova 
feedback models, suggesting that stellar and supernova feedback has little effect on the physical properties 
of the large-scale IGM. 

The observational temperature-density relation of the IGM is derived from the entire SDSS region \citep{Lim2018ApJ}, 
while the simulated relations shown here are obtained in a small volume around the Coma cluster. It is thus possible 
that the relation depends on large-scale structures, and the similarity between the non-AGN models and observation is
just a coincidence. As a check, we also present results for the G3-H-VD simulation (Table \ref{TableSimulations}).
This is a simulation for a large void in the SDSS region 
\citep[see][for more details]{Li2022ELUCID}. Interestingly, there is no significant difference 
in the median relationship between G3-H-VD, G3-H-CM, and GZ-SBnA-CM, although the fraction of high-temperature IGM is slightly
larger in G3-H-CM and GZ-SBnA-CM than in G3-H-VD. This suggests that when AGN feedback is switched off, 
the $T_{\rm IGM}$-$\rho_{\rm m}$ relation of the IGM is not strongly affected by large-scale structure, 
indicating that the simulated relations presented above may be valid for the global relation.
It is also worth noting that a more careful study, which mimics the {\it Planck} beam, nulling filter, 
CMB and infrared background subtraction, is required to properly quantify the uncertainty in the observational data.

Our findings demonstrate that when AGNs are deactivated, the simulated $T_{\rm IGM}$-$\rho_{\rm m}$ relation is 
not affected significantly by the large-scale structure, the stellar/supernova feedback model, and the method used to solve
the fluid equations. More importantly, the simulated $T_{\rm IGM}$-$\rho_{\rm m}$ relations without AGN feedback 
are in good agreement with the observations, suggesting that, over a broad range of density, 
the AGN feedback may not have a significant impact on the IGM temperature. The large discrepancy between simulations 
containing AGN feedback and the observational data suggests that the implemented AGN feedback in these 
simulations is too strong in heating the IGM.

\section{Summary and discussion}\label{sec_sum}

We conducted a series of zoom-in, constrained simulations (CSs) to investigate the formation and evolution of 
the Coma galaxy cluster and its surroundings. We employed two different galaxy formation models, \gizmo-\simba\ (GZ-SB) 
and \gadget-H (G3-H), which differ in that GZ-SB includes the growth of supermassive black holes and AGN 
feedback while G3-H does not. Our CSs, which accurately reproduce the large-scale structures of the local
universe, enable us to make a more detailed, ``one-to-one'' comparison between models and observations. To this end, we 
divided the simulated high-resolution region into two regions according to their distance to the Coma cluster center and 
compared observations and models in the two regions separately. We also included observations of large-scale gas, 
such as the ICM, CGM and IGM, to constrain and calibrate the galaxy formation model. The Coma cluster has abundant
observational data for its ICM, from the radio to the X-ray band, which provides valuable information about the evolutionary
history of the Coma ICM, but has rarely been used to constrain the subgrid physics in galaxy formation models. 
Our main results are summarized below.

Generally, the GZ-SB model is able to reproduce the observed stellar mass function in the Coma region, except 
that it significantly underestimates the population of quiescent galaxies in the low-density region. 
This discrepancy cannot be detected using the conventional approach of considering the low- and high-density regions 
together. Additionally, the GZ-SB model underestimates the stellar and ISM metallicities across the entire stellar 
mass range in both the low- and high-density regions. This suggests that some internal processes
are not modeled correctly, as environmental effects may be negligible in the low-density region. 
The metallicity of the ICM is also underestimated, indicating that the total metal yield of the fiducial \simba\ model 
is too low. To address this problem, one may weaken the AGN feedback to increase the star formation rate so as to 
produce more metals. However, this does not work for low-mass galaxies, in which AGN feedback is not important. 
It seems that an increase in the star formation rate and in supernova feedback strength is needed to fix the problem. 

Our new model assuming higher star formation rate and supernova feedback strength matches the observed 
metallicities in stars, the ISM and the ICM better than the fiducial model. The predicted \MSZR\ and its scatter are 
also in agreement with the observational results in both low- and high-density regions. In the high-density region, 
owing to strangulation, quiescent galaxies have a higher \MSZR\ than star-forming galaxies, which is consistent with 
the observations. In low-density regions, we find that the intrinsic scatter of the \MSZR\ is very small, 
indicating that these are good places to study the origin of the \MSZR . The new simulations also reproduce the 
ICM metallicity and its radial dependence. The amplitude of the predicted \MIZR\ is consistent with the observations, but 
the slope is very different. Interestingly, in the new simulation, the predicted stellar mass function (SMF) 
for quiescent galaxies in the low-density region matches the observations much better because the high star-formation 
rate and supernova feedback reduce the amount of cold gas and make galaxies more susceptible to AGN feedback.

The G3-H model is able to accurately reproduce the total SMF in both low- and high-density regions. However, when 
star-forming and quiescent galaxies are separated, large discrepancies become apparent. It predicts too many 
star-forming galaxies and too few quiescent galaxies. In particular, there are almost no quiescent galaxies in the 
low-density region, which implies that environmental effects are a major factor in quenching galaxies in the G3-H model. 
On the other hand, the predicted \MSZR\ and \MIZR\ for star-forming galaxies in the low-density region are in good agreement 
with the observational data in terms of amplitude, slope, and scatter. In the high-density region, it appears that 
the environmental effect is too strong, likely owing to the strong feedback from stars and supernovae, 
so that star-forming and quiescent galaxies share the same \MSZR , in contrast to the observational result. 
Additionally, the G3-H model under-predicts the ICM metallicity. All of these findings suggest that the 
G3-H model is successful in capturing the formation history of star-forming galaxies but fails to 
properly quench galaxies.  An additional internal quenching mechanism, such as AGN feedback, seems to be required 
for the G3-H model to work.

We analyzed the physical properties of the ICM of the Coma cluster, such as electron density, electron temperature, and entropy. 
The predictions of the simulations at $z=0$ are broadly consistent with the observational results. The GZ-SB model and 
its modifications can reproduce the temperature profile, while the G3-H model can recover the density and entropy profiles. 
We found that these properties are significantly affected by a recent merger event. Our further investigations reveal that 
the simulation results can be improved if the merger occurs slightly later, and so the discrepancy may be produced 
by the uncertainty in the reconstruction to precisely recover the highly non-linear merger event. Allowing for this uncertainty, the two simulations without AGN feedback, G3-H-CM and GZ-SBnA-CM, can accurately reproduce 
the three profiles. Simulations with AGN feedback, GZ-SB-CM, \GZSBrw /rs-CM, can accurately reproduce the temperature and 
density profiles, but overestimate the entropy profile in the inner region. This is because the heating by AGN feedback 
stabilizes the entropy in the inner region at a high level and produces a central core.

The eROSITA X-ray observation revealed a strong bow-like shock feature in the inner region and a massive subcluster 
in the southwest of the Coma cluster.  Our two non-AGN simulations, GZ-SBnA-CM and G3-H-CM,  produce similar features 
at similar locations, while all simulations with AGN feedback basically fail to recover these features. In our simulations, 
the shock is produced by the collision of the halo gas between the main Coma cluster and a subhalo.  When AGN feedback 
is implemented, the halo gas in the subhalo is severely reduced by AGN feedback prior to the collision so that 
a strong shock is not produced.  A similar situation is also found in the southwest subcluster, which has a halo mass 
of $M_{200c} = 10^{13.6}\msun$.  In GZ-SB-CM and \GZSBrw /rs-CM, strong AGN feedback expels its gas so that its X-ray emission 
becomes very weak, very different from what is seen in the non-AGN simulations and in the observation. 

We also investigated the properties of the IGM. Observational studies revealed that the IGM temperature ($T_{\rm IGM}$) 
increases with increasing cosmic density ($\rho_{\rm m}$). The two non-AGN models predict a 
$T_{\rm IGM}$-$\rho_{\rm m}$ relation that is consistent with the observations within the observational 
uncertainty. Furthermore, our tests demonstrate that the simulated $T_{\rm IGM}$-$\rho_{\rm m}$ relation is
insensitive to the adopted star formation model, hydrodynamic solver and cosmic variance. 
On the other hand, when AGN feedback is included, the IGM in low-density regions can be heated to a very high 
temperature, which is very different from what is observed.

Our study indicates that to accurately recover the stellar mass function of both star-forming and quiescent galaxies, 
an internal quenching mechanism, such as AGN feedback, must be implemented. However, we find that models without AGN feedback 
can accurately reproduce physical properties of the observed ICM and IGM, while those with strong AGN feedback usually fail.  
This implies that quenching mechanisms may only operate on relatively small scales  
and do not significantly alter the gas properties on cluster and larger scales.
Moreover, models without AGNs can also reproduce the metallicity-stellar mass relations for star-forming 
galaxies in the low-density region, indicating that AGN feedback should not significantly affect the 
properties of star-forming galaxies before quenching them.

Our study demonstrates that the ``one-to-one'' approach provided by our constrained 
simulations is a powerful tool to constrain galaxy formation models. 
Investigating low- and high-density regions separately can effectively differentiate between environmental and internal 
effects. Additionally, the ``one-to-one'' study with the combined multiple-band observational data can significantly increase 
the constraining power of observational data. For instance, considering the star, ISM and ICM metallicity together 
gives us a more comprehensive picture of star formation activity. These are particularly important for the calibration
of model parameters. The physical properties of the ICM and IGM are not significantly affected by star formation models 
or hydrodynamic solvers and thus are ideal probes of AGN feedback. Our study shows that, with a reliable constrained simulation, 
the properties and structures of the ICM within one single cluster, such as the Coma cluster,  
can put a significant constraint on the AGN feedback. Thus, constrained simulations, such as the Coma 
cluster simulations presented here, provide a unique and effective platform to test, verify, and calibrate 
galaxy formation models.

\section*{Acknowledgements}

We thank Eugene Churazov for help in using their eROSITA data.  
This work is supported by the National Natural Science Foundation of China (NSFC, Nos. 12192224, 11733004 and 11890693), CAS Project for Young Scientists in Basic Research, Grant No. YSBR-062, and the Fundamental Research Funds for the Central Universities. We acknowledge the science research grants from the China Manned Space Project with No. CMS-CSST-2021-A01, and CMS-CSST-2021-A03. WC is supported by the STFC AGP Grant ST/V000594/1 and the Atracci\'{o}n de Talento Contract no. 2020-T1/TIC-19882 granted by the Comunidad de Madrid in Spain. He also thanks the ERC: HORIZON-TMA-MSCA-SE for supporting the LACEGAL-III project with grant number 101086388 and Ministerio de Ciencia e Innovación for financial support under project grant PID2021-122603NB-C21. NK was supported by NASA ATP grant 80NSSC18K1016 and NSF grant AST-2205725. The authors gratefully acknowledge the support of Cyrus Chun Ying Tang Foundations. The work is also supported by the Supercomputer Center of University of Science and Technology of China. 

\bibliography{ref.bib}

\nolinenumbers
\clearpage

\end{document}